\documentclass[twocolumn,trackchanges,tighten]{aastex63}

\usepackage{amsmath}
\usepackage{xcolor}

\received{}
\revised{}
\accepted{}
\submitjournal{ApJ}

\shorttitle{The 3D dust and opacity distribution of protoplanets}
\shortauthors{Krapp, Kratter and Youdin}

\begin{document}

\title{The 3D dust and opacity distribution of protoplanets in multi-fluid global simulations}

\correspondingauthor{}
\email{krapp@arizona.edu}

\author[0000-0001-7671-9992]{Leonardo Krapp}
\affiliation{Department of Astronomy and Steward Observatory,  University of Arizona, Tucson, Arizona 85721, USA}

\author[0000-0001-5253-1338]{Kaitlin M. Kratter}
\affiliation{Department of Astronomy and Steward Observatory,  University of Arizona, Tucson, Arizona 85721, USA}

\author[0000-0002-3644-8726]{Andrew N. Youdin}
\affiliation{Department of Astronomy and Steward Observatory,  University of Arizona, Tucson, Arizona 85721, USA}
\affiliation{The Lunar and Planetary Laboratory, University of Arizona}

\begin{abstract}
The abundance and distribution of solids inside the Hill sphere are central to our understanding of the giant planet dichotomy.  Here, we present a three-dimensional characterization of the dust density, mass flux,  and mean opacities in the envelope of sub-thermal and super-thermal mass planets.  We simulate the dynamics of multiple dust species in a global protoplanetary disk model accounting for dust feedback.  We find that the meridional flows do not effectively stir dust grains at scales of the Bondi sphere.  Thus the dust-settling driven by the stellar gravitational potential sets the latitudinal dust density gradient within the planet envelope.   Not only does the planet's potential enhance this gradient, but also the spiral wakes serve as another source of asymmetry.  These asymmetries substantially alter the inferred mean Rosseland and Planck opacities. In cases with the moderate-to-strong dust settling, the opacity gradient can range from a few percent to more than two orders of magnitude between the mid-plane and the polar regions of the Bondi sphere.  Finally, we show that this strong latitudinal opacity gradient can introduce a transition between optically thick and thin regimes at the scales of the planet envelope.  We suggest that this transition is likely to occur when the equilibrium scale height of hundred-micron-sized particles is smaller than the Hill radius of the forming planet.  This work calls into question the adoption of a constant opacity derived from well-mixed distributions and demonstrates the need for global radiation hydrodynamics models of giant planet formation which account for dust dynamics.
\end{abstract}

\keywords{}

\section{Introduction} \label{sec:intro}

The core accretion model for giant planet formation relies on the complex interplay of gas and solids in a protoplanetary disk \citep[][]{Pollack1996, Youdin2013}.
Despite a mass fraction of only 1\% relative to the gas, the solids control much of the planet formation process. Both planetesimals and pebbles impact the growth rate of the solid core and final metallicity \citep{Ormel2010,Lambrechts2012, Alibert2018}. Small grains dominate the envelope opacity \citep[][]{Pollack1994, Piso2015} and thus the gas cooling rate in the radiative zones \citep[e.g.,][]{Podolak2003,Hubickyj2005}. In turn, the gas cooling rate controls the  growth timescale of the planetary envelope throughout the onset of runaway growth, when the envelope starts to dominate the planet mass \citep{Piso2014}.  
Eventually the envelope separates from the disk and hydrodynamic accretion is no longer limited by cooling, but by the supply of gas that becomes depleted by gap opening and disk dispersal \citep[][]{Lissauer2009, Ginzburg19}. We emphasize that, in core accretion theory, envelope cooling, and therefore dust opacities, are essential for a planet's transition to a gas giant.

In this work we present a three-dimensional multi-species characterization of the dust density, dust mass flux,  and mean opacities in the envelope of embedded protoplanets. 
The distribution of solids in the vicinity of a protoplanet is set by complex dynamical processes, including the gravity of the central star and of the planet, and drag forces with the disk and envelope gas, which are affected by flows triggered by the planet.  
Three-dimensional simulations have elucidated the hydrodynamic flows that couple the planet envelope and surrounding disk gas \citep{Tanigawa2012}.  
While local shearing-box simulations can more readily achieve high resolution within the Bondi sphere \citep[e.g.,][]{Kuwahara2020}, global simulations model planet disk interactions more accurately, e.g.\ by capturing the full orbits in the horseshoe region. 
Global models have become increasingly sophisticated, moving from pure hydrodynamics \citep[e.g.,][]{Wang2014,Fung2015}, to radiation hydrodynamics \citep[e.g.,][]{Ayliffe2009,DAngelo2013,Lambrechts2016,Szulagyi2016,Kurokawa2018,Schulik2019} and magnetohydrodynamics \cite[][]{Gressel2013}. 

These simulations have altered the  standard picture of 1D core accretion models, where low mass gas envelopes grow hydrostatically around a core as they cool \citep{Pollack1996, Piso2014}. 
In particular, three-dimensional simulations have shown that hydrodynamic flows penetrate within the Hill and Bondi radii, which limits the ability of this gas to cool \citep[][]{Ormel2015, Moldenhauer2021}. 
However, the gas that is deeper in the envelope is more slowly recycled and can still cool, if at a reduced rate \citep[][]{Cimerman2017}.  
Furthermore, radiation hydrodynamics calculations show that the strength of dust opacities\footnote{Note that we refer to ``dust opacity'' as the opacity of the gas which main source are small dust grains. Molecular opacity is usually negligible at the outer envelope for temperatures below $10^{3}\,{\rm K}$ \citep[][]{Freedman2008}. } has a significant effect on the nature of recycling flows and on envelope convection \citep[][]{Zhu2021}.  
Thus a hydrodynamic determination of 3D distributions of dust opacities -- an issue explored in this paper -- is a crucial step to a fuller understanding of how 3D flows affect the accretion of planetary envelopes. 

In this work, we move towards a self-consistent treatment of thermal physics at the planet-disk interface by exploring the size-dependent particle distribution at scales of the Bondi sphere in global, three-dimensional models. 
We use novel multi-fluid 3D simulations including dust feedback, to model the non-uniform dust distribution in the vicinity of the planet. 
We study the implications for envelope opacity and thus the thermal physics of the cooling and an accreting planet. 
Our simulations allow us to explore how the interplay between vertical settling and planet-disk interaction impacts the distribution of solids and mean opacities at the scale of the Bondi sphere of sub-thermal mass planets.
Thus, our simulations are complementary to  previous studies that have focused on the large scale meridional circulation of solids affected by gap-opening planets \citep[][]{Fouchet2007,Bi2021,Binkert2021}. In addition, these numerical simulations may inform updated boundary conditions for long-timescale,  one-dimensional core accretion models \citep[e.g.,][]{Lee2015}.

\begin{figure*}[t]
    \centering
    \includegraphics[scale=0.9]{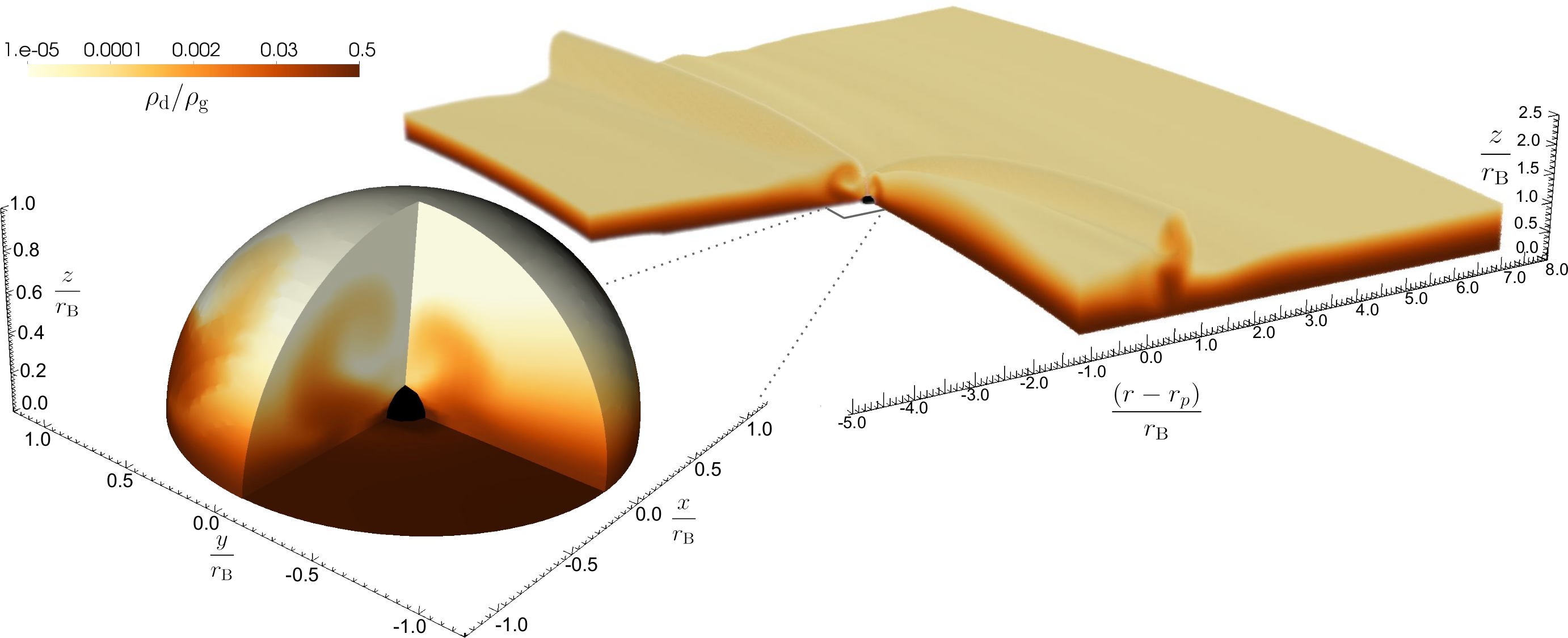}
    \includegraphics[scale=0.95]{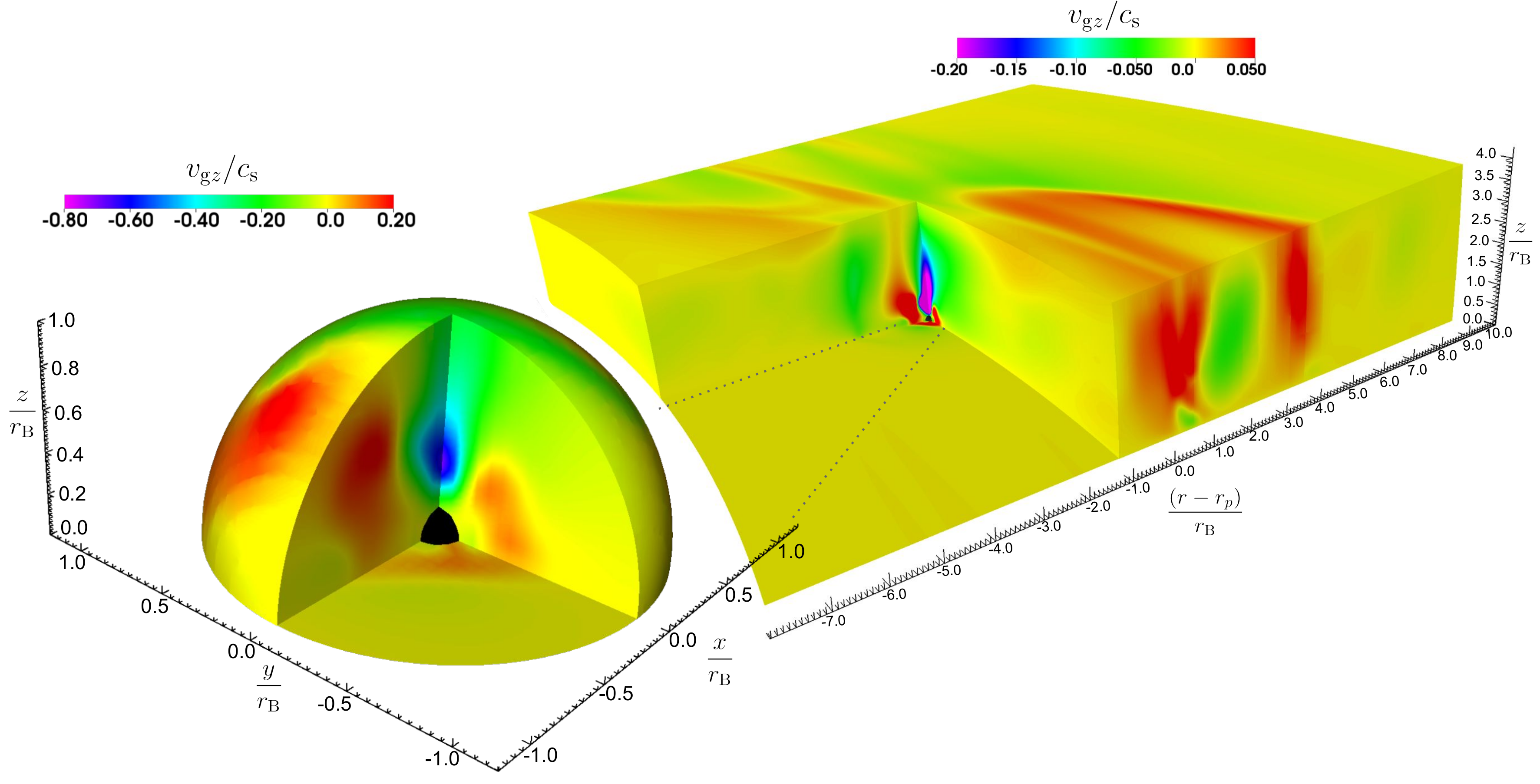}
    \caption{Top panel: Dust-to-gas density ratio for the run \texttt{SUB-d5} for the full domain, with a zoom-in at the Bondi sphere.  The dust perturbation driven by the spiral wakes extends well within the Bondi sphere, as described in Section\,\ref{sec:results_opacity}. Bottom panel: Vertical gas velocity normalized by the sound speed for the run \texttt{SUB-d5}. The horseshoe region is characterized by meridional gas outflow where $v_{{\rm g}z} \sim 0.05 c_{\rm s}$.  }
    \label{fig:fig0}
\end{figure*}

This work is organized as follows:
\paragraph{Model setup and description of the three-dimensional flow:} In Section\,\ref{sec:numerical} we present the equations, numerical method, and disk model that we adopt in this work.
In Section\,\ref{sec:results} we describe the gas and dust flows at the scale of the Bondi sphere, which are consistent with previous models that neglect feedback. 
Focusing on the dust density distribution, in Section\,\ref{sec:dust_density} we investigate the deviations from the initial settling equilibrium introduced by the planet potential, highlighting the dust-to-gas ratio anisotropies due to settling induced latitudinal variations, and spiral density wave induced azimuthal ones.

\paragraph{Opacities calculation at the Bondi sphere and Conclusions:}  In Section\,\ref{sec:opacity} we estimate how departures from spherical symmetry affect the Rosseland and Planck mean opacity at the Bondi radius. 
In Section\,\ref{sec:results_opacity} we show the existence of an anisotropic opacity distribution driven by dust settling. 
We moreover calculate the photon mean-free path and describe a latitudinal transition between optically thick and thin regimes at the Bondi sphere of a sub-thermal mass planet in Section\,\ref{sec:discussion}. We therefore identify the disk locations and conditions in which this transition may develop. 
Finally, in Section\,\ref{sec:conclusions} we summarize the main conclusions and emphasize the need for multi-species self-consistent radiative transfer disk models for future work.

\section{Numerical Method}
\label{sec:numerical}
The numerical simulations in this work are carried out using the multi-fluid code FARGO3D \citep{Benitez-Llambay2016,Benitez-Llambay2019}. 
We first introduce the equations in Section\,\ref{sec:equations}, then continue with a summary of the key features of the numerical method in Section \ref{sec-fargonum}. We defer discussion of the post-processing analysis required to produce realistic opacity maps to Section \ref{sec-opacmethod}.

\subsection{Equations}
\label{sec:equations}

Our multi-fluid hydrodynamic simulations solve the following set of equations.
First, the continuity equations are
\begin{eqnarray}
 \partial_t \rho_{\rm g} + \nabla \cdot\left( \rho_{\rm g} \mathbf{v}_{\rm g}\right) & = & 0\,, \nonumber \\
  \partial_t \rho_{{\rm d}j} + \nabla \cdot\left( \rho_{{\rm d}j} \mathbf{v}_{{\rm d}j} + {\bf j}_{{\rm d}j}\right) & = & 0\,, \label{eq:continuity}
\end{eqnarray}
where $\rho_{\rm g}$, $\rho_{{\rm d}j}$, ${\bf v}_{\rm g}$ and ${\bf v}_{{\rm d}j}$ correspond to the gas and dust densities and velocities, respectively.
The dust diffusion flux is given by
\begin{equation}
\label{eq:diffusion}
{\bf j}_{{\rm d}j} = -D \left( \rho_{\rm g} + \rho_{{\rm d}j}\right) \nabla \left(\frac{\rho_{{\rm d}j}}{\rho_{\rm g} + \rho_{{\rm d}j}}\right)
\end{equation}
where $D$ is the diffusion coefficient. 
The momentum equations for the gas and dust-species are
\begin{eqnarray}
  \partial_t \mathbf{v}_{\rm g} + \mathbf{v}_{\rm g} \cdot \nabla \mathbf{v}_{\rm g} & =& - \frac{\nabla P}{\rho_{\rm g}} - \nabla \Phi +\frac{1}{\rho_{\rm g}}\nabla \cdot {\bf \tau} - {\bf F}_{\rm g}\,,  \nonumber  \\
  \partial_t \mathbf{v}_{j} + \mathbf{v}_{j} \cdot \nabla \mathbf{v}_{j}  & = & - \nabla \Phi  - {\bf F}_{j}\,,
  \label{eq:momenta}
\end{eqnarray}
for $j=1,\dots,N$. The gas pressure is defined as $P=c^2_{\rm s}\rho_{\rm g}$, with $c_{\rm s}$ the sound speed.
The terms ${\bf F}_{\rm g}$ and ${\bf F}_{j}$ denote the accelerations due to the drag force between gas and dust-species, respectively, and are defined as 
\begin{eqnarray}
    {\bf F}_{\rm g} & = & \frac{\Omega}{\rho_{\rm g}} \sum^N_{j=1} \frac{\rho_{{\rm d}j}}{T_{{\rm s}j}} \left( {\bf v}_{\rm g} - {\bf v}_{{\rm d}j} \right)\,,\nonumber \\
    {\bf F}_{\rm j} & = & \frac{\Omega}{T_{{\rm s}j}}\left( {{\bf v}_{{\rm d}j}-\bf v}_{\rm g} \right)\,,
\end{eqnarray}
where $T_{{\rm s}j}$ corresponds to the Stokes number of the $j$-th dust species and $\Omega$ is the orbital frequency  \citep[e.g.,][]{Epstein1923,Whipple}. In this work we assume a constant Stokes number and neglect the momentum and mass transfer between dust species. 

The gravitational potential $\Phi$ includes the contributions from the central star and the planet and neglects the indirect term, thus
\begin{equation}
    \Phi = -\frac{GM_\odot}{r} - \frac{GM_{\rm p}}{\sqrt{|{\bf r} - {\bf r}_{\rm p}|^2 + r^2_s}}
\end{equation}
where  $r_s$ is a softening length used to avoid a divergence at the planet location, and $M_{\rm p}$ and ${\bf r}_{\rm p}$ correspond to planet mass and radial vector position.
The viscous stress tensor, ${\bf \tau}$, is given by
\begin{equation}
    {\bf \tau} = \rho_{\rm g} \nu \left( \nabla {\bf v} + \left(\nabla {\bf v}\right)^{T} - \frac{2}{3} (\nabla \cdot {\bf v}) {\bf 1} \right)
\end{equation}
with $\nu$ the gas viscosity. 
%
\begin{table*}[]
	\begin{center}
    \caption{List of Numerical Simulations }
    \begin{tabular}{lcccccccc}
    	\hline
    	\hline
    Run  & $\epsilon$ & $\delta$ & $\alpha$ &  $M_{\rm p}/M_{\odot}$ & $\Delta \phi$ & $N_{\phi} \times N_{r} \times N_{\theta}$ & $N_{\rm dust}$ & $T_{\rm s, min} - T_{\rm s, max}$ \\
    \decimals
    \hline
      \texttt{SUB-d4} & 0.01 & $5\times10^{-4}$ & $2\times10^{-4}$ & $2.6\times10^{-5}$ & $2\pi$ &$4800\times192\times36$ & $11$ & $2.3\times10^{-6} - 10^{-2}$\\
      \texttt{SUB-d5} & 0.01 & $5\times10^{-5}$ & $2\times10^{-5}$ & $2.6\times10^{-5}$ & $2\pi$ &$4800\times192\times36$ & $11$ & $2.3\times10^{-6} - 10^{-2}$ \\
      \texttt{SUB-d6} & 0.01 & $5\times10^{-6}$ & $2\times10^{-6}$ & $2.6\times10^{-5}$ & $2\pi$ &$4800\times192\times36$ & $11$ & $2.3\times10^{-6} - 10^{-2}$ \\
      \texttt{SUP-d5} & 0.01 & $5\times10^{-5}$ & $2\times10^{-5}$ & $7.8\times10^{-5}$ & $2\pi$ &$4800\times192\times36$ & $11$ & $2.3\times10^{-6} - 10^{-2}$  \\   
      \texttt{SUB-d5-half-sig0} & 0.01 & $5\times10^{-5}$ &$2\times10^{-5}$ & $2.6\times10^{-5}$ & $\pi$ &$2400\times192\times36$ & $11$ & $2.3\times10^{-5} - 10^{-1}$\\
      \texttt{SUB-d5-e1} & 0.1 & $5\times10^{-5}$ &$2\times10^{-5}$ & $2.6\times10^{-5}$ & $2\pi$ &$4800\times192\times36$ & $11$ & $2.3\times10^{-6} - 10^{-2}$\\
     \texttt{SUB-d5-half} & 0.01  & $5\times10^{-5}$ &$2\times10^{-5}$ & $2.6\times10^{-5}$ & $\pi$ &$2400\times192\times36$ & $11$ & $6.3\times10^{-6} - 10^{-2}$\\
       \texttt{SUB-d5-half-cnv} & 0.01  & $5\times10^{-5}$ &$2\times10^{-5}$ & $2.6\times10^{-5}$ & $\pi$ &$4800\times384\times72$ & $5$ & $6.3\times10^{-6} - 10^{-2}$\\
      \texttt{SUB-d5-half-adia} & 0.01 & $5\times10^{-5}$ &$2\times10^{-5}$ & $2.6\times10^{-5}$ & $\pi$ &$4800\times384\times72$ & $5$ & $6.3\times10^{-6} - 10^{-2}$\\
      \texttt{SUP-nd} & -- & -- &$2\times10^{-5}$ & $7.8\times10^{-5}$ & $2\pi$ &$4800\times192\times36$ & $\rm NO$ & -- \\
     \texttt{SUB-nd} & -- & -- &$2\times10^{-5}$ & $2.6\times10^{-5}$ & $2\pi$ &$4800\times192\times36$ & $\rm NO$ & --\\
    \hline
	\end{tabular}
    \label{tab:runs}
	\end{center}
	\tablecomments{\footnotesize The symbols are: $\epsilon$ is the total dust-to-gas mass ratio, $\delta$ is the settling diffusion equilibrium parameter, $\alpha$ is the effective viscosity parameter, $\Delta \phi$ is the extension of the azimuthal domain, $N_{\phi}$, $N_{r}$ and $N_{\theta}$ number of cells in azimuth, radius and latitude, respectively. The last column corresponds to the minimum and maximum Stokes numbers of the dust distribution. }
\end{table*}

\subsection{ $\rm{FARGO3D}$ numerical simulations} \label{sec-fargonum}

We solve equations \ref{eq:continuity} and \ref{eq:momenta} using an improved version of the code FARGO3D with a more efficient velocity update method, which also allows for effective parallelization of multiple fluids \citep[][]{Krapp2020}. 
We perform global three-dimensional simulations on a spherical mesh centered at the star, with coordinates $(r,\phi,\theta)$.  
The numerical domain comprises only half of the disk in the vertical direction, with $\theta_{\rm min} = \pi/2-0.1$ and $r\in[0.4,2.1]$. 
Results will also be described as a function of spherical coordinates centered at the planet defined as $(\tilde{r}, \tilde{\phi}, \tilde{\theta})$, which is a coordinate transformation rather than the simulated domain.

We use a non-uniform static mesh in $r$ and $\theta$, defined to provide adequate resolution at the Hill sphere and close to the mid-plane where dust settles. 
The grid is obtained using the grid density function $\psi(s) = a/s + \xi/((s-s_0)^2+\xi^2)$  \citep{Benitez-Llambay2018}. 
For the radial coordinate $s=r$, $s_0=r_{\rm p}$, $a=1$ and $\xi = 0.05$, whereas for the vertical polar coordinate we set $s=\theta$, $s_0=\pi/2$, $a=0$ and $\xi = 0.02$. 
This configuration provides about $16-20$ cells at the Bondi radius for the $\verb|SUB|$ runs in the radial and vertical direction with $N_{r}\times N_{z} = 192\times36$. 

Unlike $r$ and $\theta$, the grid is uniform in the azimuthal direction, which is required for this implementation of the FARGO scheme for orbital advection \citep{Masset2000}. To achieve a resolution at the Bondi radius comparable to that of the radial and vertical direction we set $N_{\phi}=4800$. 
In all the runs the softening length $r_{s}$ is set to approximately two cells in the azimuthal direction. 
A numerical convergence study of our simulations is presented in Appendix \ref{appendix:convergence}. 
\begin{figure*}[]
    \centering
    \includegraphics[scale=0.525]{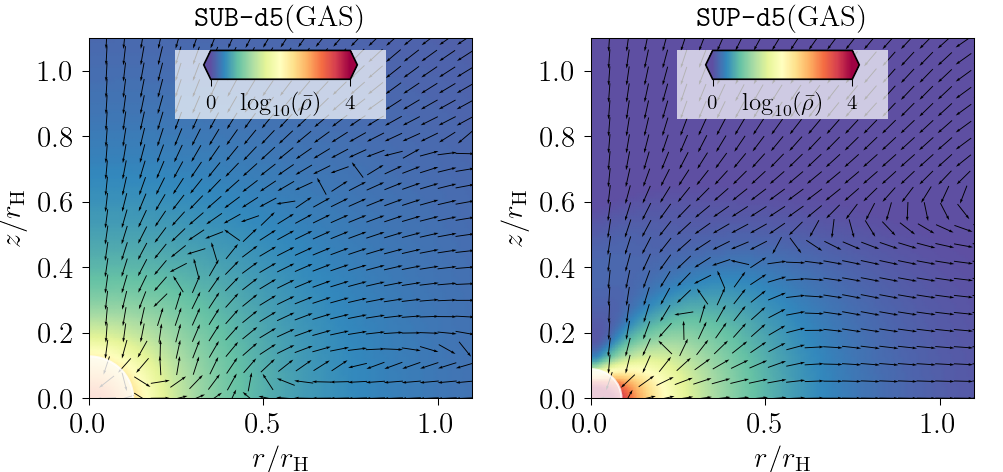}
    \includegraphics[scale=0.525]{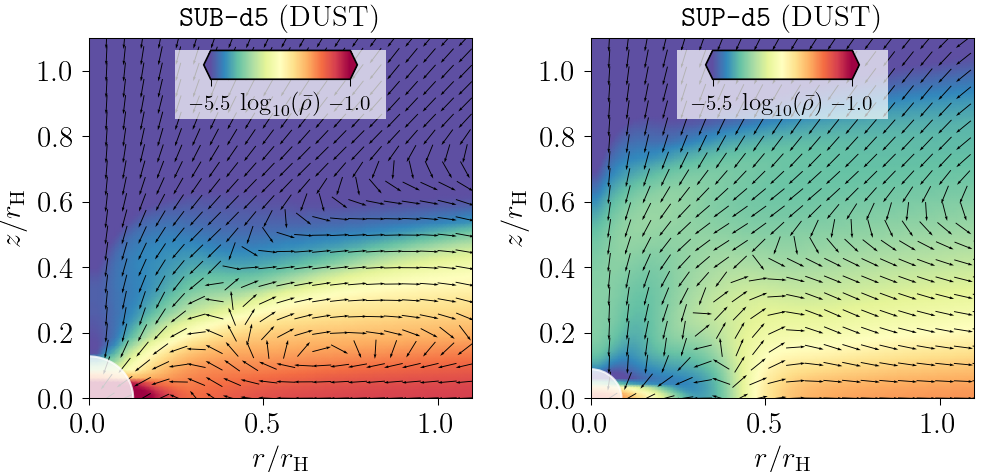}
    \caption{Azimuthally averaged (along $\tilde{\phi}$) meridional  flow pattern for the runs \texttt{SUB-d5} and \texttt{SUP-d5}. Left panels correspond to the gas flow whereas right panels show the results for dust with Stokes number $T_{\rm s}=0.01$ (the most decoupled species in this work). Gas density is normalized by the initial density, whereas dust density is normalized by gas density. }
    \label{fig:fig_meridional}
\end{figure*} 

\bigskip
\bigskip
\bigskip

\subsubsection{Boundary conditions}

At the inner and outer disk radius we employ reflecting boundary conditions for $v_{{\rm g}r}$ and $v_{{\rm d}r}$. 
The gas density and azimuthal velocity are extrapolated to match the initial conditions, whereas for the dust, the radial derivatives of the density and dust azimuthal velocity are set to zero. 
We additionally include wave-damping buffer zones that restore the density to the initial value while preventing undesired reflections that may perturb the flow \citep[][]{DeVal-Borro2006}.

At the upper boundary, the gas density and azimuthal velocity are extrapolated to match the initial conditions, while for the dust $\partial_{\theta} v_{{\rm d}\phi} = \partial_{\theta} \rho_{\rm d} = 0$. 
A reflecting boundary is adopted for $v_{{\rm g}\theta}$ and $v_{{\rm d}\theta}$. 
Since we simulate only one disk hemisphere, we also adopt reflecting boundary conditions at the disk equator for $v_{{\rm g}\theta}$ and $v_{{\rm d}\theta}$, otherwise, the latitudinal derivative is set to zero.
The initial conditions and parameters of all simulations are described in Section\,\ref{sec:disk} and Table\,\ref{tab:runs}.

\subsection{Disk Model}
\label{sec:disk}

Our base disk model is adapted from Appendix A of \cite{Masset_2016}. 
We adopt as a reference distance to the central star the planet semi-major axis, $r_p = r_0 = 1$ and typically quote time in units of the inverse of the orbital frequency $\Omega_0 = \sqrt{G M_{\odot}/r_0^3}$.
The gas sound speed and disk aspect ratio are defined as
\begin{equation}
    c_{\rm s} = c_{{\rm s}0}\, \left(\frac{r}{r_0}\right)^{-\beta/2}\,, \quad \quad h = c_{\rm s}v^{-1}_{\rm K}\,,
\end{equation}
where $v_{\rm K} = \sqrt{GM_{\odot}/r}$ is the Keplerian velocity. 
We set $\beta=1$ to simulate a disk with constant aspect ratio, that is a gas scale-height $H = hr$. We fix the disk aspect ratio to $h=0.035$ in all our runs.
The gas density is given by
\begin{equation}
    \rho_{\rm g} = \frac{\Sigma_0}{\sqrt{2\pi} h r_0} \left(\frac{r}{r_0}\right)^{-\sigma-1} \displaystyle{\sin(\theta)^{(-\beta-\sigma-1 \, + 1/h^2)}}
\end{equation}
with $\sigma = 1/2$ and the surface density $\Sigma_0 = 6.366197 \times 10^{-4} M_{\odot}/r^2_0$, which corresponds to $\Sigma_0 \simeq 200\, {\rm g}\,\rm{cm}^{-2}$ at $r_0 = 5.2 \rm{AU}$.
The radial and vertical velocity are set to zero, wheres the azimuthal velocity is 
\begin{equation}
    v_{{\rm g}\phi} = v_{\rm K} \sqrt{1-(\beta+\sigma+1) h^2} = v_{\rm K} \sqrt{1- 2.5 h^2}
\end{equation}
The gas viscosity corresponds to $\nu = \alpha c^2_{\rm s}/\Omega_{\rm K}$, with $\Omega_{\rm K}=v_{\rm K}/r$ the orbital Keplerian frequency. The values of $\alpha$ adopted for each run are shown in Table\,\ref{tab:runs}. 

The dust azimuthal velocity is initialized with a Keplerian rotation profile; although this neglects the size-dependent impact of gas drag, the velocities quickly re-adjust under the influence of the sub-Keplerian gas. The dust density of the $j$-th species is given by
\begin{equation}
    \rho_{{\rm d}j} = \frac{\epsilon_j\Sigma_0}{\sqrt{2\pi} h_{{\rm d}j} r_0 } \left(\frac{r}{r_0}\right)^{-\sigma-1} \displaystyle{\sin(\theta)}^{(-\beta-\sigma-1\, + 1/h^2_{{\rm d}})}
\end{equation}
where 
\begin{equation}
    \epsilon_j = \epsilon \frac{T^{4-s}_{{\rm s} j+1} - T^{4-s}_{{\rm s} j}}{T^{4-s}_{{\rm s, max}} - T^{4-s}_{{\rm s, min}}}\,,
    \label{eq:distribution}
\end{equation}
where $\epsilon$ is the total dust-to-gas mass ratio in the simulated domain. 
We fix $s=3.5$ and consider dust-species with $T_{\rm s, max} = 0.01$, which corresponds to nearly $\rm{cm}$-size particles at the Hill radius assuming $r_0=5.2 \rm{AU}$ in the adopted disk model. 
The aspect-ratio of the $j$th dust-species, $h_{{\rm d}j}$, is obtained from the settling and diffusion equilibrium of dust particles 
\citep[e.g.,][]{Dubrulle1995},
\begin{equation}
\label{eq:h_d}
    h_{{\rm d}j} \equiv h\,\sqrt{\frac{\delta}{\delta + T_{{\rm s}j}}}\,
\end{equation}
where $\delta$ is a free parameter that sets the scale-height of the dust, $H_{{\rm d}j} \equiv h_{{\rm d}j} r$, from the balance between settling (due to vertical stellar gravitational acceleration) and the diffusion flux defined in Eq.\,\ref{eq:diffusion}.
The diffusion coefficient $D = \delta c^2_{\rm s}/\Omega$; the value of $\delta$ is shown in (see Table\,\ref{tab:runs}) for each run.

Note that in absence of a self-consistent mechanism that sustains particles above the mid-plane, the choice of $\delta$ is unconstrained.
However, in viscous disk models meant to approximate isotropic turbulent diffusion, it is usually assumed that $\delta$ is of the order of the effective viscosity parameter $\alpha$. 
Moreover, since the diffusion coefficient satisfies $D \sim \nu$ for $T_{\rm s} \ll 1$, we safely neglect the size dependency of the diffusion coefficient\citep[e.g.,][]{YoudinLithwik2007}. 
Therefore, the adopted initial settling equilibrium may slightly deviate from that obtained with self-consistent stirring at scales $z\simeq H$ \citep[see e.g.,][]{Fromang2009}. The dust distributions are fixed for each run and do not account for coagulation, collisions, or evaporation of larger solids, which would also alter the opacity and size distributions.

\subsection{Planet Model}
Most of our analysis focuses on the gas and dust dynamics at the Hill and Bondi spheres, with radius $r_{\rm H} = r_{p} \left( M_p/3M_{\odot} \right)^{1/3}$ and $r_{\rm B} = G M_p/c^2_{\rm s}$, respectively. We refer to the gas on these scales as part of the planet's envelope, regardless of whether the gas is bound to the planet or cycling through.
Because our simulations are locally isothermal, these definitions are fixed throughout each run.
We consider only two planet masses (see Table\,\ref{tab:runs}), and denote our simulations with the prefix $\texttt{SUB}$ (sub-thermal) when $M_{p}/M_{\rm th} = r_{\rm B}/ H \simeq 0.6$, where $M_{\rm th} = h^{3} M_{\odot}$ is the disk thermal mass. 
Equivalently, the prefix $\texttt{SUP}$ is used accordingly for runs with $M_p/M_{\rm th} = r_{\rm B}/ H  \simeq 1.8$ (super-thermal). 
The planet is implemented as a potential only, meaning that the initial conditions within the Hill sphere are identical to that in the disk.
We do not consider any accretion sink for the planet, that is we do not remove mass and momentum inside the smoothing length.  
The resolution of our 3D global simulations is adequate to capture the dynamics at the interface between the planet envelope and the protoplanetary disk. 
This interface is traditionally defined as $r_{\rm envelope} = {\rm min}(r_{\rm B}, r_{\rm H})$, although it is a reference location, not a sharp barrier.

\begin{table*}[]
\begin{center}
		\caption{Simulations diagnostics at the Hill radius.
		\label{table:results1}}
\begin{tabular}{lccccccccccc}
	\decimals
	\hline
	\hline
Run  & $\langle \epsilon^{\rm mid}/\epsilon^{\rm zero} \rangle_{\tilde{\phi}}$ &$\mathcal{R}_{5.2}$ &  $W_{5.2}$ & $\mathcal{R}_{30}$ & $W_{30}$ &  $\dot{M}_{\rm g}^{\rm net}$  & $\dot{M}_{\rm g}^{\rm in}$  & $\dot{M}_{\rm g}^{\rm out}$ &  $\dot{M}_{\rm d}^{\rm net}$  & $\dot{M}_{\rm d}^{\rm in}$ & $\dot{M}_{\rm d}^{\rm out}$ \\
\hline
\texttt{SUB-d5}  &  96.2  &  2.7  &  0.93  &  287.0  &  0.18  &  -0.061  &  6.982  &  7.043  &  0.023  &  0.141  &  0.117 \\
$\texttt{SUP-d5}$  &  49.6  &  2.3  &  0.82  &  87.8  &  0.15  &  0.926  &  19.735  &  18.810  &  0.007  &  0.231  &  0.224 \\
$\texttt{SUB-d4}$  &  8.9  &  1.4  &  0.96  &  9.3  &  0.93  &  -0.054  &  6.744  &  6.798  &  0.023  &  0.117  &  0.094 \\
$\texttt{SUB-d6}$  &  534.9  &  13.7  &  0.81  &  937.7  &  0.06  &  -0.071  &  7.001  &  7.072  &  0.024  &  0.156  &  0.132 \\
$\texttt{SUB-d5-half-sig0}$  &  663.5  &  12.4  &  0.88  &  2595.6  &  0.04  &  -0.011  &  0.706  &  0.717  &  0.009  &  0.017  &  0.007 \\
$\texttt{SUB-d5-e1}$  &  53.5  &  2.3  &  0.94  &  169.9  &  0.21  &  -0.230  &  6.978  &  7.208  &  0.060  &  1.507  &  1.447 \\
$\texttt{SUB-d5-half}$  &  60.1  &  2.4  &  0.93  &  147.9  &  0.31  &  -0.058  &  6.803  &  6.861  &  0.020  &  0.121  &  0.101 \\
$\texttt{SUB-d5-half-conv}$  &  170.6  &  4.4  &  0.78  &  564.7  &  0.13  &  -0.321  &  7.020  &  7.340  &  0.036  &  0.146  &  0.110 \\
$\texttt{SUB-d5-half-adia}$  &  36.9  &  2.8  &  0.92  &  63.4  &  0.38  &  -0.020  &  5.760  &  5.780  &  0.022  &  0.136  &  0.114 \\
\hline
  \end{tabular}
\end{center}
\tablecomments{\footnotesize Simulation diagnostics are obtained at $20$ and $40$ planet orbits for sub-thermal (\texttt{SUB-}) and super-thermal (\texttt{SUP-}) mass  planets, respectively.  $\mathcal{R}$is the mid-plane to pole opacity ratio, while $W$ represents the ratio of the shell-averaged opacity to that for a well mixed distribution (defined in Section \ref{sec-opacmethod}). Note that the subscripts 5.2 and 30 refer to the radial distance in AU (the location of Jupiter and Neptune in our Solar System, respectively) chosen to evaluate $\mathcal{R}$ and $W$. The estimated mass fluxes $\dot{M}_{\rm g}$ and $\dot{M}_{{\rm d}}$ correspond to the time averaged values (from the first two orbits until the final integration time) across the Hill radius. The mass fluxes values are normalized to $10^{-5} M_{\rm jup} {\rm yr}^{-1}$.} 
\end{table*}

\section{Gas and Dust flow} \label{sec:results}
Our simulations demonstrate that at scales of the envelope, the dust flow, and thus dust-to-gas density ratio, is inherently three dimensional and decouples from the gas dynamics, meaning that dust is not well described by a well-mixed fluid with a fixed $\epsilon$. 
The stellar vertical gravity, spiral wakes, and the planet-potential drive strong variations in the dust-to-gas density ratio at scales of the Bondi sphere. The meridional circulation within the planet envelope is too weak to loft dust grains above the equilibrium scale-height set by the balance of vertical gravity and diffusion. 
The dust distribution deviates from the settling equilibrium only where the spiral wakes intersect the Bondi sphere. 
We furthermore find that the multi-species dust feedback has minimal impact on the meridional flow patterns and overall gas mass flux cross the Hill surface for a total dust-to-gas mass ratio $\epsilon \leq 0.1$.

We focus on the outcome from runs \texttt{SUB-d4}, \texttt{SUB-d5} and \texttt{SUB-d6}.
These simulations have a planet mass of $M_{\rm p} = 2.6\times10^{-5} M_{\odot} \simeq 0.6 M_{\rm th}$ and only differ by the strength of diffusion and therefore the dust scale-height relative to the Hill radius. 
For comparison, we also include the run \texttt{SUP-d5} with a planet mass of $M_{\rm p} = 7.8\times10^{-5} M_{\odot} \simeq 1.8 M_{\rm th}$. 
Since our simulations feature near-thermal mass planets, we focus our discussion on calculations at the Hill semi-sphere. Note that for all the \texttt{SUB}-runs  $r_{\rm H} \simeq r_{\rm B}$.

\begin{figure*}[t]
    \centering
    \includegraphics[scale=0.9]{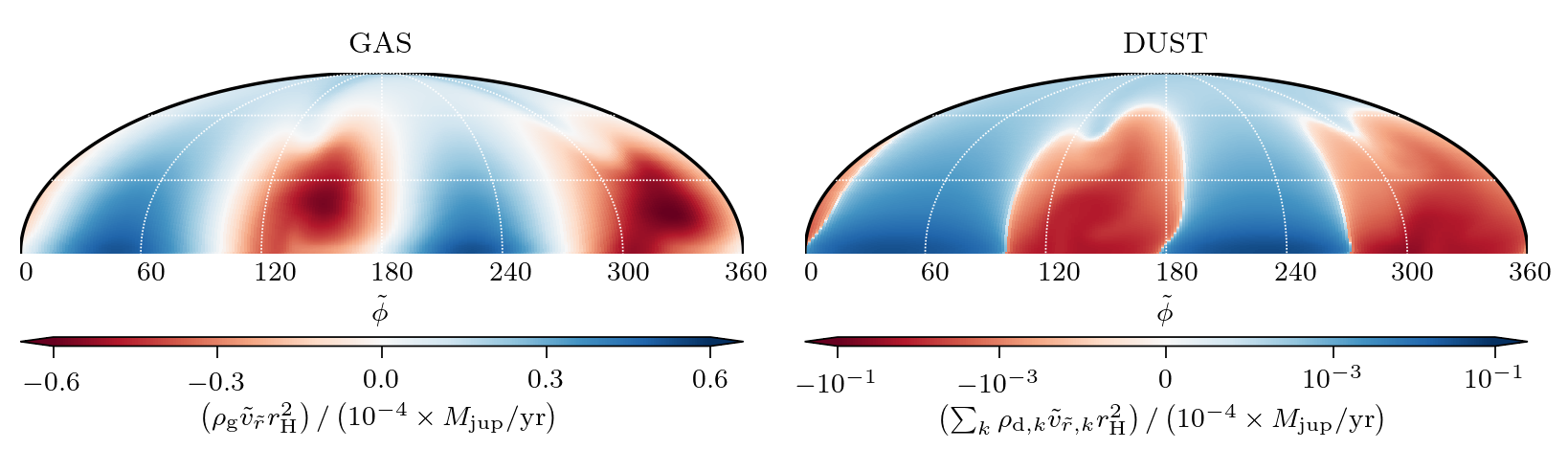}
    \caption{Inflow (blue) and outflow (red) regions for the gas and the dust for the simulation \texttt{SUB-d5} at the Hill ($\sim$ Bondi) semi-sphere. The mass flux is characterized by an inflow at the polar regions that extends down to the mid-plane. Strong outflow close to the mid-plane is carried by the spiral wakes highlighted in Fig.\,\ref{fig:fig0}. The inner spiral wake intersects Bondi sphere at $\tilde{\phi} \sim 120$, while the outer does it at $\tilde{\phi} \sim 300$. }
    \label{fig:fig3}
\end{figure*}
\begin{figure*}[]
    \centering
    \includegraphics[]{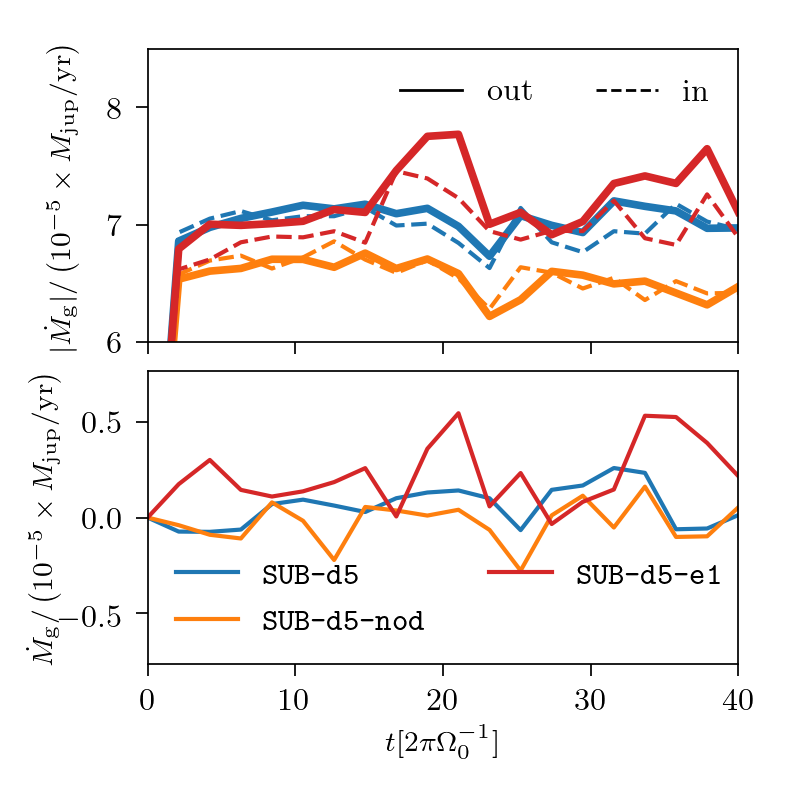}\hfill
    \includegraphics[]{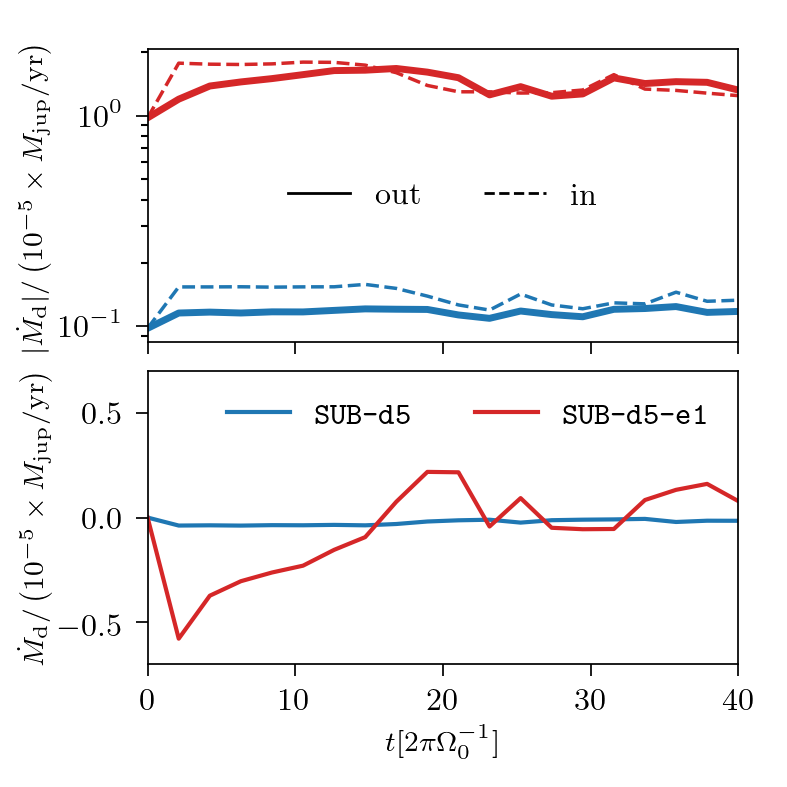}
    \caption{Shell integrated mass flux at the Hill semi-sphere for the runs \texttt{SUB-d5} (blue), \texttt{SUB-d5-e1} (red) and \texttt{SUB-d5-nod} (orange) as a function of time.
    The top panels show the values for inflow with solid lines and outflow with dashed lines, whereas the bottom panels show the net flux.
    The left and right panels correspond to the gas and dust, respectively.
    The flux at the Hill sphere of all simulations seems to have reached a steady-state with a net mass flux ${\dot M}_{\rm g} \simeq 10^{-7} M_{\rm jup}/{\rm yr}$ and ${\dot M}_{\rm d} \simeq 10^{-8} M_{\rm jup}/{\rm yr}$. 
    Increasing the total dust-to-gas mass ratio from $0.01$ to $0.1$ does not alter the net gas mass flux, although the imbalance between inflow and outflow can be of the order of $\sim 10$\%.}
    \label{fig:mass_flux_time}
\end{figure*}

\subsection{Overview of three-dimensional morphology}

We begin with a qualitative summary of the 3D morphology that characterizes our runs.
In Fig\,.\ref{fig:fig0} we show a snapshot of the run \texttt{SUB-d5} at $t=20 \times 2\pi\Omega^{-1}_0$ that includes the total dust density (top panel) and the vertical gas velocity (bottom panel).
Qualitatively similar flow patterns are obtained at the scales of the Hill radius and at the horseshoe region in all our runs.
Dust grains excited by the planet's spiral wakes are lofted above their initial scale-height $H_{{\rm d}j}$. 
The lifted stream of dust seems to have a vertical turn-over  about $z \simeq H_{\rm g}$, where $v_{{\rm g}z}$ seems to change sign. 
The zoomed-in plots  correspond to the Bondi semi-sphere, and shows that the perturbation introduced by the spiral wakes inside the planet envelope, disturb the dust flow down to planetary smoothing length.

In the bottom panel of Fig\,.\ref{fig:fig0} we show the gas vertical velocity for run \texttt{SUB-d5}.
While the gas vertical velocity is subsonic everywhere in the disk, the Mach number, $v_{{\rm g}z}/c_{\rm s}$, approaches unity inside the Bondi sphere, at the polar region near the smoothing length of the planet potential. 
This strong inflow is also seen in the dust species, with a vertical velocity that slightly deviates from that of the gas, consistent with the moderate-to-strong coupling. 
Mach numbers in regions with uplift ($v_{{\rm g}z}>0$) can reach values of $v_{{\rm g}z}/c_{\rm s} \sim 0.05$ near the horseshoe region, whereas $v_{{\rm g}z}/c_{\rm s} \sim 0.2$ inside the Bondi sphere. 

Comparing the zoomed-in top and bottom panel of Fig.\,\ref{fig:fig0} reveals that regions with gas $v_{{\rm g}z}>0$  are traced by an enhancement of the dust density.
Dust grains are lifted at these locations, generating the anisotropic dust-gas mixture, that deviates not only from the commonly assumed well-mixed composition, but also that of a equilibrium settled disk model \citep[see e.g.,][]{Chachan2021}.

\subsection{Meridional flow pattern}
\label{sec:meridional}

Recycling of material between the disk and the envelope may be crucial for understanding planet growth, and in particular may slow atmospheric cooling and contraction \citep[e.g.,][]{Ormel2015}. 
Even with our moderate resolution inside the Hill sphere, we still observe this characteristic meridional flow pattern. Note that the in appendix\,\ref{appendix:convergence} we directly compare our meridional flow pattern with previous works, and evaluate the numerical convergence.

In this section we focus on runs \texttt{SUB-d5} and \texttt{SUP-d5} to compare flow patterns with fixed $\delta$ across the transition from sub to super-thermal masses, where we might expect rapid gap opening to commence.

In Fig.\,\ref{fig:fig_meridional} we show the azimuthal average of the gas and dust density (only $T_{\rm s}=0.01$, as it is the most decoupled species in our runs).
On top of the mean density we plot the azimuthal average of the meridional velocity field, $\tilde{v}_{\rm mer} = (\tilde{v}_{\tilde{r}}, \tilde{v}_{\tilde{z}} )$, where the $\tilde{v}_{\tilde{r}}$ and $\tilde{v}_{\tilde{z}}$ correspond to the radial and vertical velocities obtained in a cylindrical coordinate system centered on the planet.

Our results for the gas meridional velocity resemble the characteristic flow described in both local \citep[e.g.,][]{Bethune2019} and global \citep[e.g.,][]{Fung2019} isothermal simulations, where inflow occurs mainly at the polar regions, while outflows are prominent at the mid-plane. 
We find that at the disk mid-plane, both inflow and outflow occur (see Section,\ref{sec:mass_flux}) and therefore the azimuthal average provides an incomplete description of the flow at $\theta = \pi/2$. 

The gas density for the sub-thermal mass planet remains nearly symmetric, while the super-thermal mass case shows hints of rotational flattening. However, these density structures are close enough to the smoothing length that they are resolution dependent (see Appendix\,\ref{appendix:convergence}).

Overall, the dust meridional flow follows the gas since there is a moderate-to-strong coupling. However, regions with strong gas vertical infall are depleted of dust (see Section \ref{sec:mass_flux}, Figure \ref{fig:fig3}). 
Therefore the transport of solids towards the planetary core is dominated by mid-plane flows.

For the lower mass case, the average dust layer thickness seen in  Fig.\,\ref{fig:fig_meridional} is similar to the equilibrium $H_{\rm dj}$ far from the planet.  The higher mass planet stirs dust a bit more, unsurprisingly.  
The higher mass case also shows a dust gap at $0.25 \lesssim r/r_{\rm H} \lesssim 0.5$, but this small scale feature could be resolution dependent, and needs further study to confirm.

\subsection{Mass flux at the Hill radius}
\label{sec:mass_flux}

Our simulations do not permit accretion onto the planet since we adopt a softened gravitational potential with no envelope self-gravity, nor removal of mass and angular momentum. 
Thus, we cannot explicitly measure envelope or core growth.
Instead we calculate the gas mass flux at the Hill semi-sphere to characterize the balance between inflow and outflow and assess the steady-state nature of our simulations. 
We compare our results with previous simulations to determine if the dust back reaction onto the gas significantly affects the gas mass flux.

The mass flux is estimated from the output density and velocity fields and integrated at the Hill semi-sphere as
\begin{equation}
    \dot{M} = \int^{2\pi}_{0} \int^{0}_{\pi/2} \rho ({\bf v} \cdot \hat{\tilde{r}}) r^2_{\rm H} \sin(\tilde{\theta}){\rm d}\tilde{\theta}\,{\rm d}\tilde{\phi}
\end{equation}
The values obtained for the net flux $\dot{M}_{\rm net}$, as well as inflow (${\bf v} \cdot \hat{\tilde{r}} > 0$), $\dot{M}_{\rm in}$, and outflow (${\bf v} \cdot \hat{\tilde{r}} < 0$), $\dot{M}_{\rm out}$, are shown in Table\,\ref{table:results1}.
All the values are normalized to $10^{-5} M_{\rm jup}/{\rm yr}$ and obtained after time-averaging between the $t=2 \times 2\pi \Omega^{-1}_0$ and the final integration time of each run.

In Fig.\,\ref{fig:fig3} we illustrate the mass flux through the Hill sphere for both the gas and the dust at $r_{\rm H}$. We also show in Fig\,\ref{fig:mass_flux_time} the time evolution of $\dot{M}_{\rm in}$, $\dot{M}_{\rm out}$ and $\dot{M}_{\rm net}$ for several runs.
For the gas, we find that $\dot{M}_{\rm in}$ and $\dot{M}_{\rm out}$ are typically $\sim 7 \times 10^{-5} M_{\rm jup}/{\rm yr}$ and both fluxes balance within a few percent, which is highlighted by the substantially smaller net flux, typically $\lesssim 10^{-6} M_{\rm jup}/{\rm yr}$.
The minimal net mass flux is consistent with the observed quasi-steady state and near hydrostatic balance within the Bondi sphere (see Appendix\,\ref{appendix:convergence}).
Note that for the super-thermal mass case the inflow/outflow values increase to $\sim 2 \times 10^{-4} M_{\rm jup}/{\rm yr}$.
As shown in Table\,\ref{table:results1}, this is a  $\sim 3$-fold increase between the mass fluxes for run \texttt{SUP-d5} compared to \texttt{SUB-d5}, which correlates linearly with the planet mass.  

Similar inflow and outflow distribution at $r = r_{\rm B}$ has been reported by \cite{Cimerman2017} although their simulations were local and included radiative transfer.
The measured fluxes are also in reasonable agreement with those from the previous works of \cite{Bate2003} and \cite{Ayliffe2009} for the isothermal runs with an accreting planet with mass $M_p = 10 M_{\earth}$, which is the closest case in units of the thermal mass to our work. 
Higher resolution may narrow the balance to $\lesssim 1 \%$ between $\dot{M}_{\rm in}$ and $\dot{M}_{\rm out}$ as reported for sub-thermal mass planet simulations \citep{Fung2015,Fung2019}.

Figure \ref{fig:fig3} reveals that the dust shows a similar flow pattern to the gas, albeit with significantly smaller fluxes. 
While a reduction is expected simply due to lower dust-to-gas density ratios, $\dot{M}_{\rm d} \neq \epsilon \dot{M_{\rm g}}$.
At latitudes near the mid-plane, the total dust mass flux is roughly $\sim 10$ times smaller than the gas mass flux,for $\epsilon = 0.01$. 
At the poles, $\dot{M}_d < \epsilon \dot{M}_g$, with inflow reduced to  $\dot{M}_{\rm in}\ll 10^{-8} M_{\rm jup}/{\rm yr}$.
This trend is expected due to dust settling, and is enhanced by the fact that the dust mass distribution is tilted towards the largest stokes numbers which settle efficiently.

In addition to spatial variation in the mass flux ratios, the integrated dust mass flux is in disagreement with that expected for a well-mixed distribution, i.e. 1\% of the gas (see Table\,\ref{table:results1}). 
Moreover, the net dust mass flux is always towards the planet (inflow) even though the net gas mass flux is predominately an outflow.
The difference between gas and dust mass fluxes can be attributed to fact that the dust mass flux is dominated by the species with the largest size ($\sim \rm cm$) and therefore more decoupled, therefore dust recycling may be less efficient. 

To better understand the impact of dust feedback on the gas dynamics, in Fig\,\ref{fig:mass_flux_time} we also include a run with no dust, \texttt{SUB-nod} and a run with a total dust-to-gas mass ratio of 10\%, instead of the fiducial 1\%.  
We find that while feedback does modify the gas velocity locally, especially at the mid-plane where the dust-to-gas ratio approaches unity, overall the gas mass flux, mean meridional circulation, and envelope hydrostatic equilibrium seems to be agnostic to the presence of the dust at scales of $r \sim r_{\rm H}$ when $\epsilon = 0.01$.

The simulation with a larger dust-to-gas mass ratio (10\%) shows a stronger gas inflow/outflow imbalance, with a mean mass flux $\sim 4$ times larger than the \texttt{SUB-d5} case.
Although the net dust mass flux is not significantly larger than the runs with $\epsilon = 0.01$, a larger dispersion is shown in the left panel of Fig\,\ref{fig:mass_flux_time} (in comparison with the 1\% standard case) which suggests that longer integration times may be needed to reach equilibrium.
Future numerical simulations with higher resolution at the scales on which circumplanetary disks form ($r\lesssim r_{\rm H}$) will better determine if dust feedback remains negligible in the inner regions of the planet envelope, where larger concentrations of dust are expected.
\begin{figure*}[t]
    \centering
    \includegraphics[]{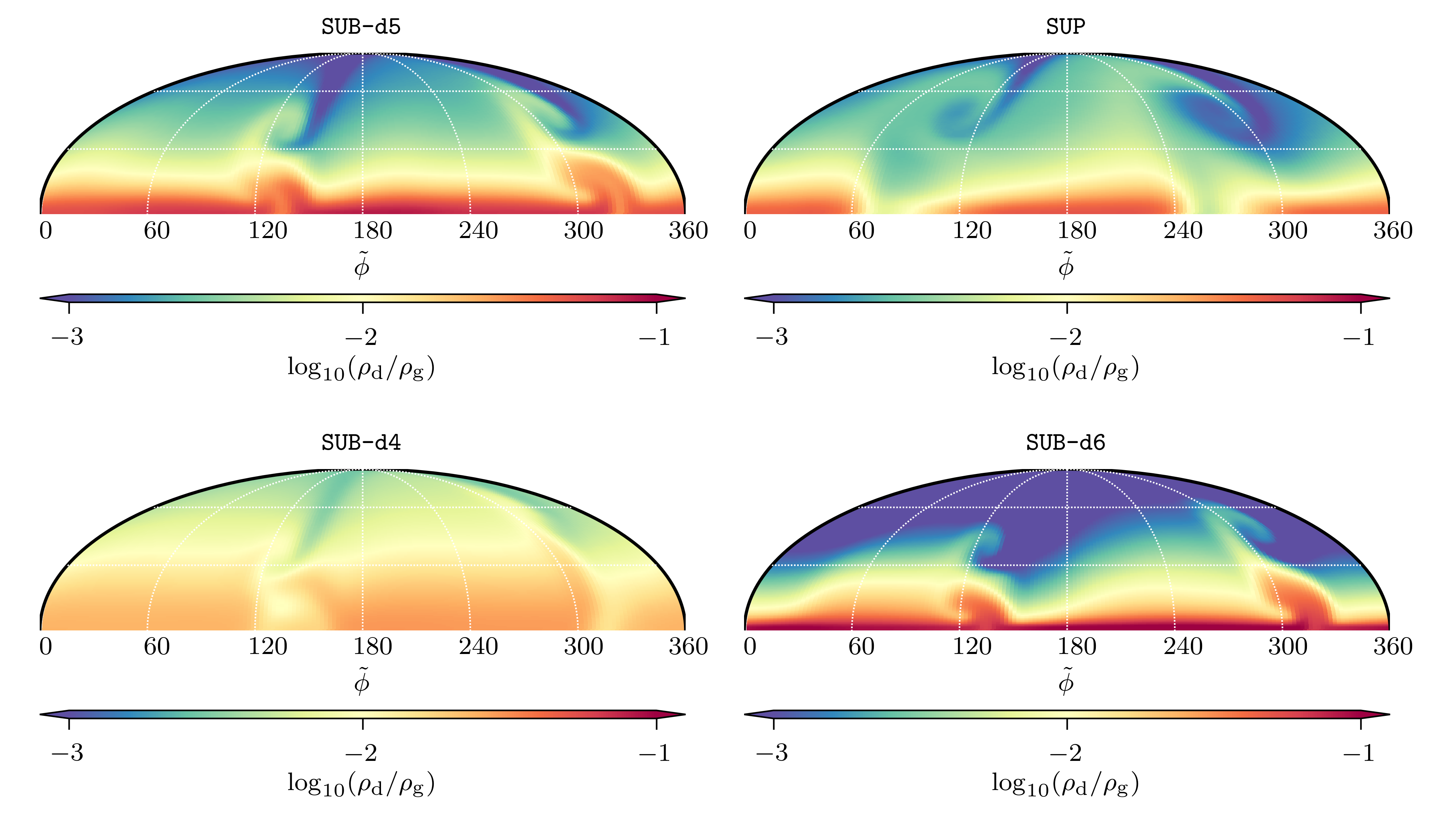}
    \caption{Dust-to-gas density ratio at the Hill sphere for the runs \texttt{SUB}, \texttt{SUP}, \texttt{SUB-d1} and \texttt{SUB-e1}. In all the runs there is a moderate anisotropic density distribution in the azimuthal direction induced by the planet's spiral wake. The planet introduces a stronger latitudinal gradient in comparison with the diffusion-settling equilibrium away from the spirals wakes perturbations.}
    \label{fig:fig1}
\end{figure*}
\begin{figure}[]
    \centering
    \includegraphics[scale=0.85]{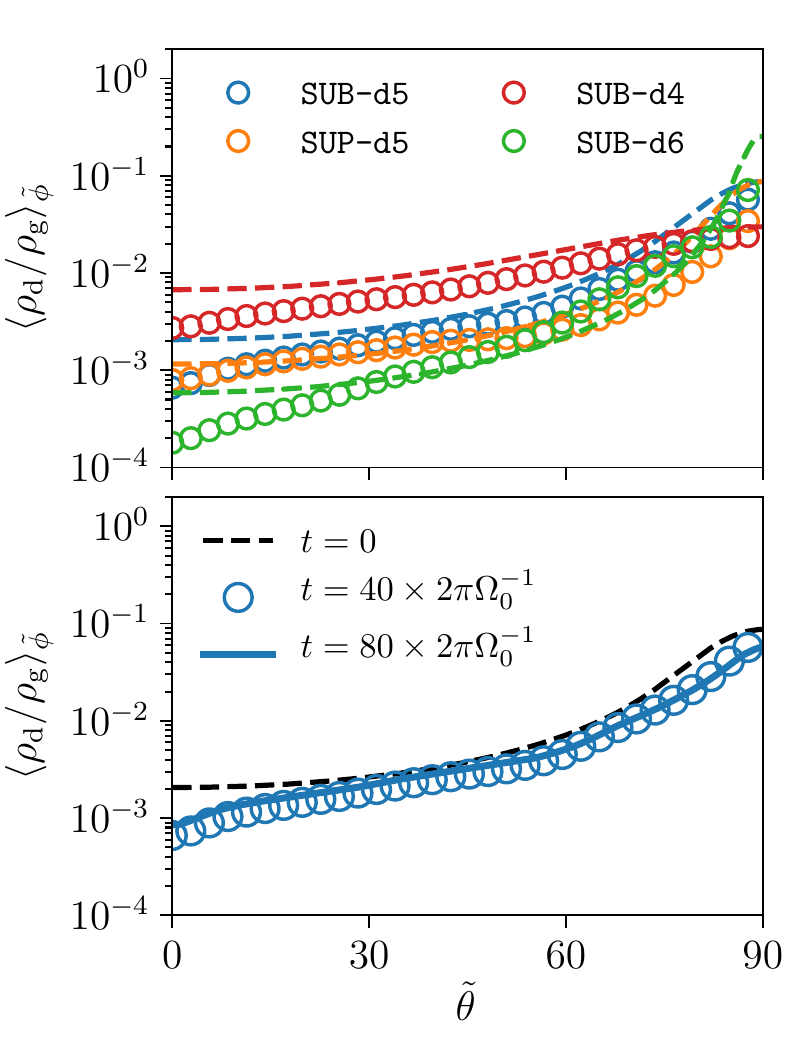}
    \caption{Total dust-to-gas mass ratio at the Hill semi-sphere, after taking the azimuthal average, from pole ($\tilde{\theta} = 0$) to equator. 
    The top panel shows the results for the runs displayed in Fig.\,\ref{fig:fig1}. Dashed lines correspond to the initial condition whereas circles were obtained at $t=20 \times 2\pi\Omega^{-1}_0$ for the run \texttt{SUP-d5} and $t=40 \times 2\pi\Omega^{-1}_0$ for the rest of the runs. 
    The bottom panel shows a comparison between different orbital times for the run \texttt{SUB-d5} which highlight the nearly steady-state dust-to-gas density ratio at the Hill sphere. }
    \label{fig:fig_average_all}
\end{figure}

\subsection{Dust density at the Hill surface}
\label{sec:dust_density}

The distribution of small solids plays a major role in shaping the opacity of the planet envelope. However, it is usually assumed that grains are all well-mixed with the gas. We, therefore, assess the validity of such approximation by describing the dust distribution obtained from our numerical simulations at the Hill hemisphere.
We find that the planet's potential steepens the latitudinal gradient in dust-to-gas ratio that is driven by the balance between settling and diffusion.
Moreover, the planet interaction with the dusty-fluids induces an anisotropic density distribution along the longitudinal coordinate, $\tilde{\phi}$. 

In Fig\,\ref{fig:fig1} we show the total dust-to-gas density ratio at the Hill radius.
Each panel has the label of the run as shown in Table\,\ref{tab:runs}.
As expected from  dust settling, the largest values of dust-to-gas mass ratio are obtained at the mid-plane.  Comparing the dust-to-gas mass ratios of runs $\texttt{SUB-d5}$, $\texttt{SUB-d4}$ and $\texttt{SUB-d6}$, it is clear that changing the diffusion parameter, which changes $H_{{\rm d}j}$ (see eq. \ref{eq:h_d}), primarily modulates the dust density gradient along the latitudinal direction a the Hill radius.

Fig.\,\ref{fig:fig1} also shows that the anisotropic dust density distribution develops at longitudes of $\tilde{\phi} \sim 120^{\circ}$ and $\tilde{\phi} \sim 300^{\circ}$ for sub-thermal mass planets.
These dominant features are ubiquitous in all runs and seem to be driven at the location of the spiral wakes excited by the planet.
The one at $\tilde{\phi}\sim 120^{\circ}$ corresponds to the inner wake, whereas the one at $\tilde{\phi}\sim 300^{\circ}$ corresponds to the outer one. 
Note that for the case of a super-thermal mass planets (therefore stronger velocity perturbations), these features are extended and shifted by a full $60^{\circ}$.
Dust density increases along these wakes, where dust is being lifted from the mid-plane well above the the dust scale height $H_{{\rm d}j}$.
However, the effective $\epsilon \lesssim 0.01$ in these wakes, and therefore dust feedback is negligible.

To facilitate a quantitative comparison between the results obtained for different $\delta$ we show in the top panel of Fig.\,\ref{fig:fig_average_all} the azimuthal average of the dust-to-gas density ratio at the Hill radius. 
Moreover, we include the dust-to-gas mass ratio at $t=0$ (dashed lines) to demonstrate that the planet potential plays a minor role in shaping the dust's latitudinal distribution at the Hill radius. 

The latitudinal deviation introduced by the planet is highlighted by the comparing the mean profiles. A clear depletion develops at the polar regions while mid-plane values are slightly enhanced. 
This deviation becomes stronger with smaller dust scale-heights. 
Finally, in the bottom panel of Fig.\,\ref{fig:fig_average_all}  we show the azimuthal average of the dust-to-gas mass ratios at different orbital times for the run \texttt{SUB-d5} to emphasize that the dust distribution at the Hill sphere has reached a steady-state. 
Typically, we find a steady-state distribution at the Hill sphere after 2-5 orbits and the flow persists in steady-state at least up to 80 orbits, the time at which we terminate our runs. 

\section{Mean Opacity} \label{sec:opacity}
We have shown in Section \ref{sec:results} that the coupled gas and multi-species dust dynamics generates substantial departures from the standard simplification of spherically symmetric, well-mixed dust.
We now translate our more realistic dust distributions into  opacities, to better infer its impact on radiative transfer in the envelope, and thus planetary cooling rates. 
We focus our calculations on the values of the opacity at the Hill radius, and examine the contributions of vertical settling and planet-disk interaction on the opacity.
We show that the planet envelope can have a latitudinal transition between optically thin and optically thick regimes, at certain locations in the disk, primarily driven by dust settling. 

\subsection{Methods: Opacity Calculations} \label{sec-opacmethod}

The mean opacities per gram of dust are obtained using the publicly available DSHARP opacity module; the details of the assumed composition, and absorption and scattering opacities calculation, are described in \cite{Birnstiel_2018}. 
To calculate the opacity per gram of total material we multiply the opacities per gram of dust by the total dust-to-gas mass ratio\footnote{The results are not altered if we instead multiply by the dust mass fraction, since $\rho_{{\rm d}} \ll \rho_{{\rm g}}$ in the cases discussed (e.g., see Fig.\,\ref{fig:fig_average_all}}. 
We denote the Rosseland and Planck mean opacity (per gram of material) as  $\bar{\kappa}_{\rm R}$ and $\bar{\kappa}_{\rm P}$, respectively. 
We calculate both opacities using the dust density from the numerical simulations from Section\,\ref{sec:results} (based on the disk model described in Section\,\ref{sec:disk}).

We will compare the values of $\bar{\kappa}_{\rm R}$ and $\bar{\kappa}_{\rm P}$  with equivalent opacities obtained assuming a well mixed dust distribution. We denote these opacities as $\bar{\kappa}_{\rm R, mix}(T)$ and $\bar{\kappa}_{\rm P, mix}(T)$.
Both are obtained assuming that each grid cell has a dust density equal to that of Eq.\,\ref{eq:distribution} times the gas density. 
Therefore, gradients in $\bar{\kappa}_{\rm R, mix}(T)$ and $\bar{\kappa}_{\rm P, mix}(T)$ are only sensitive to gradients in the gas density, and not the dust density.
Thus, these opacities are expected to be nearly constant in space and time since the gas density is in quasi-hydrostatic equilibrium at the Bondi radius (see Fig.\,\ref{fig:fig_appendix_a}), and remains symmetric at the Bondi radius. 

Since our simulations do not include radiative transfer and planet luminosity, we proceed by assuming a temperature, $T$, given by the condition $T^4 \equiv T^4_{\rm disk} + L/(16\pi\sigma_{\rm SB} \tilde{r}^2)$, where 
$L$ corresponds to the accretion luminosity of the planet, $L = GM^2_p/(R_{\rm planet} t_{\rm double})$. 
We assume a mass-doubling time $t_{\rm double}=10^{5}\,{\rm yr}$ and the planet radius, $R_{\rm planet}$, is obtained assuming a mean density of $\rho_{\rm planet} = 3 ~\rm{g}~\rm{cm}^{-3}$. 
This approximation for the temperature is valid in optically thin regimes, which may not apply inside the Bondi sphere in our simulations. However, we are focusing on the the outer regions of the envelope ($r \sim r_{\rm H}$), where $T \sim T_{\rm disk}$. Thus the choice of specific doubling time and corresponding planetary luminosity also has a minimal impact at the scales on which which we focus.

The opacities per gram of dust (and therefore the Rosseland and Planck mean opacities) depend on the total monochromatic opacity, $\kappa_\nu^\mathrm{tot}$.
To calculate $\kappa_\nu^\mathrm{tot}$ using the DSHARP module, we first interpolate the dust-density from our simulations to a discretized size-distribution with 200 bins that span from $a_{\rm min} = 10^{-5} \,{\rm cm}$ to $a_{\rm max} = 10^2 \,{\rm cm}$.
This interpolation facilitates the calculations since it arranges the simulated dust distribution to match the size-domain of the opacity in the DSHARP module.

Since we only calculate the opacity for specifics shells inside the Hill sphere, the conversion from Stokes number to particle size is done assuming a mean density, $\bar{\rho}_{\rm g} = \langle\langle \rho_{\rm g}\rangle\rangle_{\tilde{\phi}, \tilde{\theta}}$, and mean sound speed $\bar{c}_{\rm s} = \langle\langle c_{\rm s}\rangle\rangle_{\tilde{\phi}, \tilde{\theta}}$. 
This approximation is justified because the gas density is nearly constant at a fixed radius given the local hydrostatic equilibrium inside the Hill sphere, as we show in appendix\, \ref{appendix:convergence}.

Thus, for the $j$-th dust-species, the grain size is
\begin{equation}
\label{eq:size}
a_j \equiv T_{{\rm s}j} \, \sqrt{8/\pi} \bar{\rho}_{\rm g}/\rho_{\rm solid}\bar{c}_{\rm s} / \Omega(r_p)
\end{equation}
where $\rho_{\rm solid} = 1.6686\, \rm{g}~\rm{cm}^{-3}$, and $\Omega(r_p)$ the orbital frequency at the planet radius. The dust density is set to zero for those bins with maximum (minimum) size obtained from Eq.\,\ref{eq:size} that are smaller (larger) than the limits set by $a_{\rm max}$ ($a_{\rm min}$). 
The Stokes numbers included in our simulations correspond to sizes in the range of $1 \mu {\rm m} \lesssim a_j \lesssim 1 {\rm cm}$ at $r_0 = 5.2 {\rm AU}$ at the Bondi radius.
With the new interpolated distribution, we estimate the total extinction and absorption coefficients as follows
\begin{equation}
    \label{eq:extinction}
    \kappa_\nu^{\rm tot} = \frac{\sum^{N=200}_{j=1} \rho_{\rm d}(a_j)\,\kappa_\nu(a_j)  }{ \sum^{N=200}_{j=1} \rho_{\rm d}(a_j) }\,.
\end{equation}
where $\kappa_\nu (a)$ is obtained from the DSHARP module and it corresponds to $\kappa^{\rm abs}_\nu$ and $\kappa^{\rm ext}_\nu$ for the Planck and Rosseland mean opacity, respectively \citep[see e.g.,][]{Birnstiel_2018}.  
\begin{figure*}[t]
    \centering
    \includegraphics[]{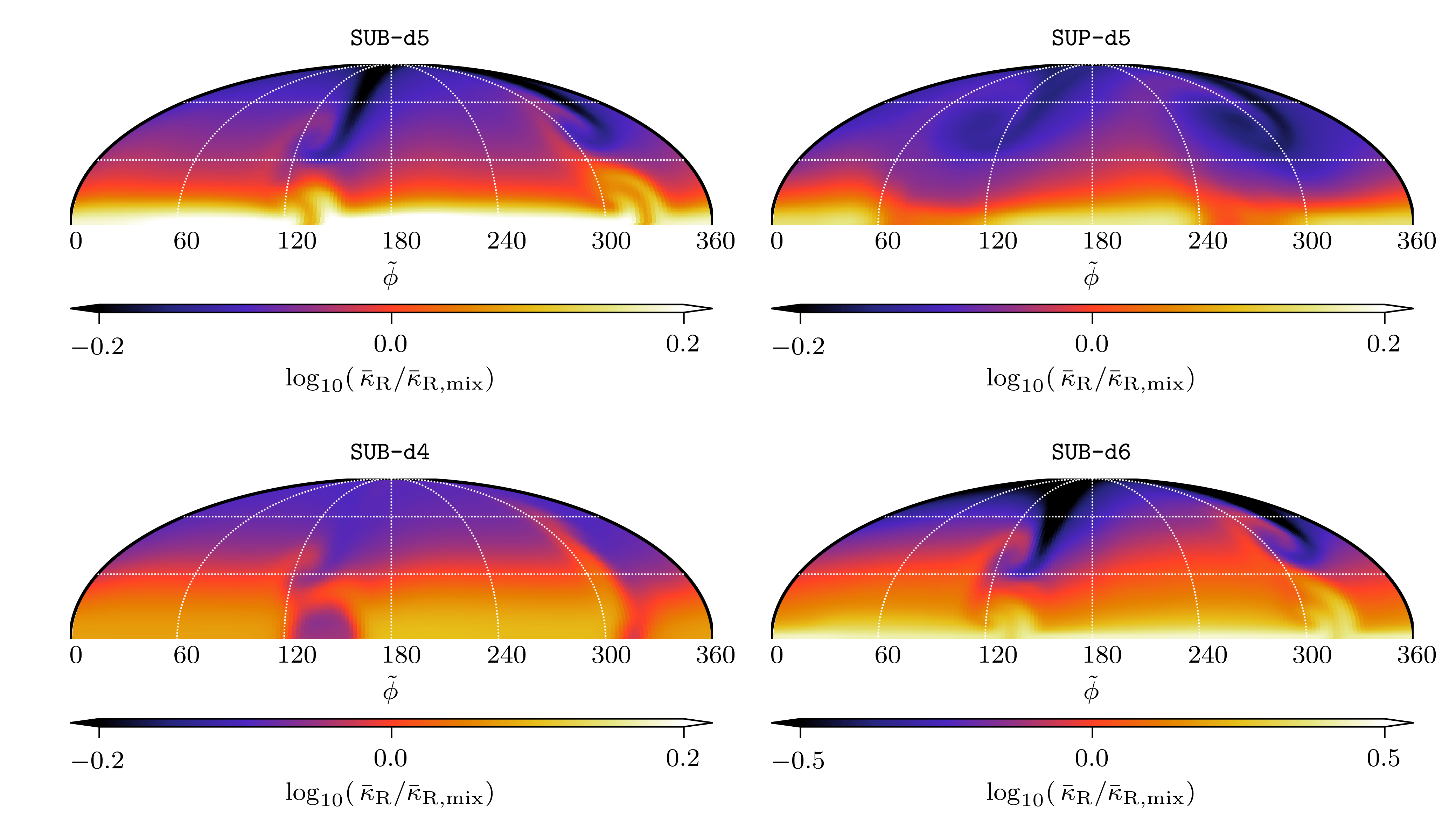}
    \caption{Rosseland Mean Opacity at the Hill semi-sphere for the runs \texttt{SUB}, \texttt{SUP}, \texttt{SUB-d1} and \texttt{SUB-e1}. All cases show a moderate anisotropic opacity from the mid-plane to the pole. The opacities were obtained by combining the simulated dust distributions and the DSHARP opacity module. The value of the Rosseland mean opacity for the well-mixed distribution is $\kappa_{\rm R,mix} \simeq 0.27 {\rm g}^{-1}{\rm cm}^{2}$ and $\kappa_{\rm R,mix} \simeq 0.37 {\rm g}^{-1}{\rm cm}^{2}$ for the runs \texttt{SUB-} and \texttt{SUP-d5}, respectively. }
    \label{fig:fig_opacity1}
\end{figure*} 

\subsection{Rosseland and Planck mean opacity}
\label{sec:results_opacity}
In this section, we present the calculations of the Rosseland and Planck mean opacity (per gram of total material) at the surface of the Hill sphere.
Based on the outcome of our numerical simulations, we show that both opacities, $\bar{\kappa}_{\rm R}(T)$ and $\bar{\kappa}_{\rm P}(T)$, deviate from $\bar{\kappa}_{\rm R/P, mix}(T)$. 
The latter correspond to opacities of a well-mixed grain distribution, that is a fixed dust-gas density ratio.

In Fig.\,\ref{fig:fig_opacity1} we show the Rosseland mean opacity at the Hill radius for the runs \texttt{SUB-d5}, \texttt{SUP-d5}, \texttt{SUB-d4} and \texttt{SUB-d6}. 
Overall, the distribution of $\bar{\kappa}_{\rm R}$ is traced by the total dust-to-gas mass ratio shown in Fig.\,\ref{fig:fig1}. 
In all cases, the mean opacity decreases from the mid-plane to the polar regions of the Hill sphere.
At the locations of the spiral wakes, where dust grains are significantly elevated compared to the dust scale-height $H_{{\rm d}j}$ expected from settling-diffusion equilibrium, the Rosseland mean opacity also increases, creating anistropy in the azimuthal direction as well.

Interestingly, the opacity distribution is similar for sub and super thermal mass planets (comparing $2.6 \times 10 ^{-5} M_{\odot}$ to $7.8 \times 10 ^{-5} M_{\odot}$) as can be seen when comparing the top left and right panel of Fig.\,\ref{fig:fig_opacity1}. 
Note however, that our results are obtained during the initial phase of gap formation for the run \texttt{SUP-d5}, and therefore we anticipate that  gap formation may affect the opacity distribution at the Hill sphere on longer time scales, because gaps may preferentially filter certain grain sizes \citep{Weber2018}.

Fig\,\ref{fig:fig_opacity1} also demonstrates that the opacity strongly depends on  settling:\footnote{assuming grain growth has proceeding beyond the initial ISM sub-$\mu$m sizes at this point in the planet formation process}  smaller values of $\delta$ generate more deviation of $\bar{\kappa}_{\rm R}$ from  $\bar{\kappa}_{\rm R, mix}$. 
We recall here that the opacity per gram of material is the product between opacity per gram of dust and total dust-to-gas mass ratio.
The opacity per-gram-of-dust (whose bulk value is dominated by the smaller grains) is nearly constant at latitudes $\tilde{\theta} \lesssim 60^{\circ}$, however, strong settling produces larger latitudinal gradients in the dust-to-gas mass ratio.
Thus, the dust-to-gas density ratio plays a crucial role in shaping the opacity gradient. 
\begin{figure}[]
    \centering
    \includegraphics[]{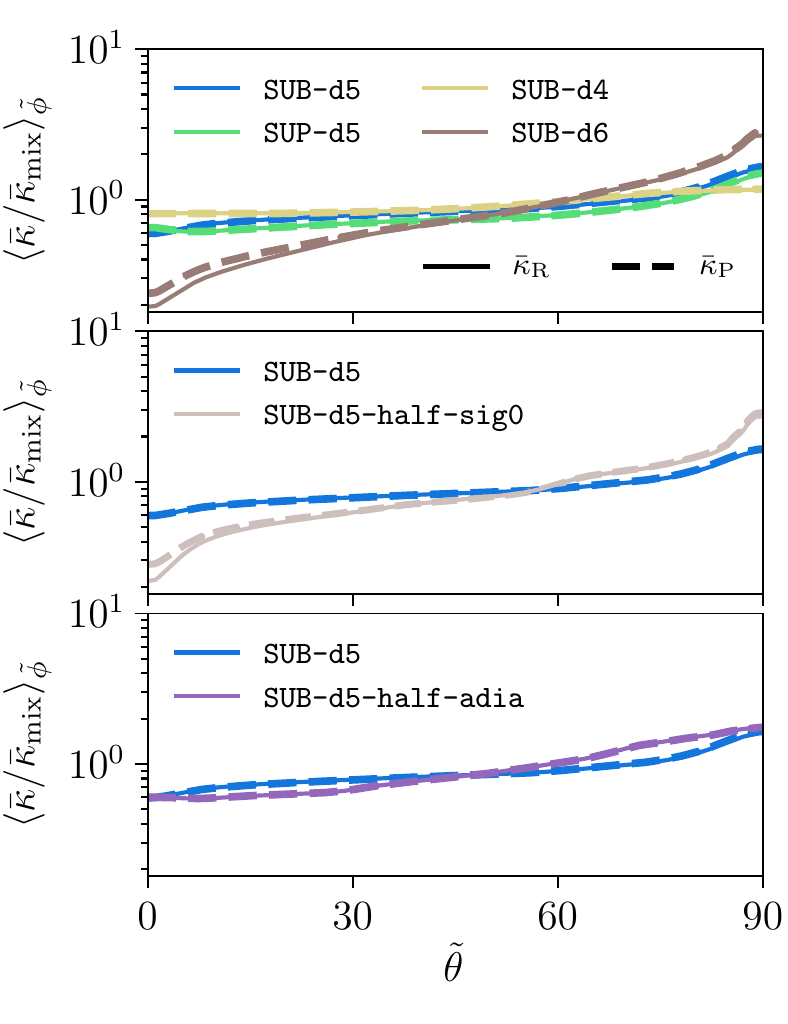}
    \caption{Azimuthal average of the Rosseland and Planck mean opacity at the Hill semi-sphere.
    We normalize the opacity by $\kappa_{\rm mix}$ which corresponds to the Planck or Rosseland mean opacities of a well-mixed size distribution.
    The opacities are estimated from the simulated data at $t=20 \times 2\pi\Omega^{-1}_0$ for the run \texttt{SUP-d5} and $t=40 \times 2\pi\Omega^{-1}_0$ for the rest of runs. 
    Strong settling ($\delta \lesssim 10^{-5}$) introduces a gradient in the opacity with respect to a well-mixed distribution. The gradients are also characterized by the mid-plane to pole ratios, $\mathcal{R}$, and the spherically integrated inverse opacity, $W$, (see Table\,\ref{table:results1}).
    }
    \label{fig:kappa_average}
\end{figure}

To better compare the effect of dust settling on the opacity distribution, we show in Fig.\,\ref{fig:kappa_average} the azimuthal average of the Rosseland mean opacity at the Hill sphere. %
All values were obtained assuming parameters from the disk model described in Section\,\ref{sec:disk} with the planet's orbit at $r=5.2 \rm{AU}$.
We also indicate the Planck mean opacity with dashed lines.
The results shown in the top panel indicate that deviations between $\bar{\kappa}_{\rm R}$ and $\bar{\kappa}_{\rm R, mix}$ are minor for $\delta = 5\times10^{-4}$, while  reaching a factor a few for $\delta = 5\times10^{-5}$. 
When $\delta = 10^{-6}$, we find an order of magnitude discrepancy between the mid-plane and polar regions.  

The latitutidinal opacity gradients are stronger for lower disk surface densities, such as those expected in the outer disk.
Therefore, the deviations from a well-mixed distribution are even more pronounced for a planet forming beyond $10$ au. To assess this case, we consider a planet forming at a location with surface density $\Sigma_0 = 20\, {\rm g}{\rm cm}^{-2}$. 
This is shown in the middle panel of Fig.\,\ref{fig:kappa_average}. 
Since dust settling is the major driver of the opacity gradient, and the dust scale-height for a given particle size depends on the gas density and local sound speed, lower density regions favor the presence of smaller grains close to the mid-plane. 

\begin{figure}
    \centering
    \includegraphics[scale=0.9]{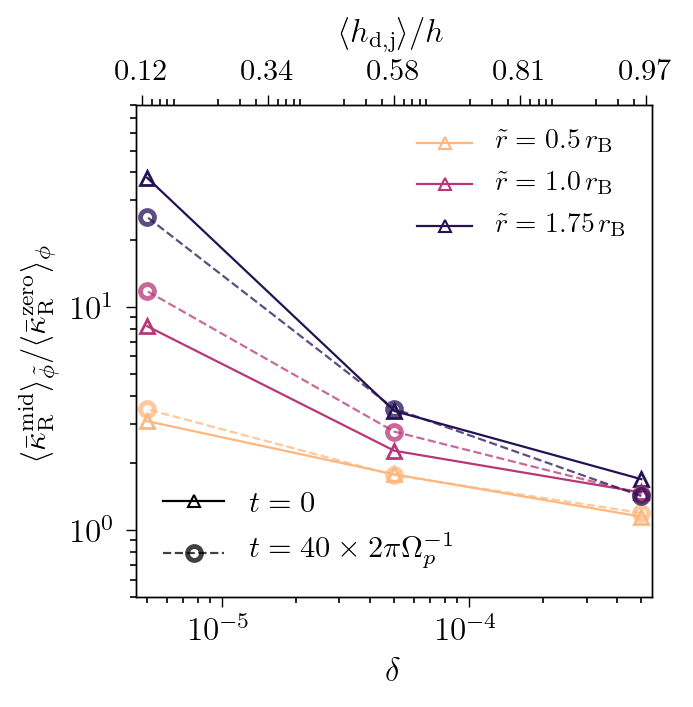}
    \caption{ Equator-to-pole ratio of Rosseland mean opacity for the runs \texttt{SUB-d6}, \texttt{SUB-d5} and \texttt{SUB-d4} after 40 planet orbits compared with the initial ratio given by the dust settling. Overall, the interplay between dusty-fluids and the planet potential only introduces variations of order unity on the azimuthal average of the Rosseland mean opacity. Our 3D PPDs multi-fluid simulations  suggest that settling due to the stellar gravitational acceleration is the main driver of the envelope opacity gradient at scales of the Bondi radius when $M_p \sim M_{\rm th}$. }
    \label{fig:fig_compare_settling}
\end{figure}

Fig.\,\ref{fig:fig_average_all} shows that the mean dust density profile at $r_{\rm B}$ deviates little from the initial settling equilibrium due to perturbations from the planet. To quantify how this profile impacts opacity, in Fig.\,\ref{fig:fig_compare_settling} we show  the ratio between the azimuthal average at the pole and mid-plane: $\langle \kappa_{{\rm R,} 5.2}\rangle_{\tilde{\phi}}^{\rm mid} /\langle\kappa_{{\rm R,} 5.2}\rangle_{\tilde{\phi}}^{\rm zero} \equiv \mathcal{R}_{5.2}$. 
To quantify the impact of the planet compared to dust settling alone, values are shown for $t=0$ and after the envelope has reached steady state. $\mathcal{R}$ is evaluated for a shell at the Bondi semi-sphere as well as shells  at $\tilde{r} = 0.5 r_{\rm B}$ and $\tilde{r} = h r = 1.75 r_{\rm B}$, all assuming a planet at $r=5.2{\rm AU}$ from the central star. 
In all cases displayed in Fig\,\ref{fig:fig_compare_settling}, the ratio $\mathcal{R}_{5.2}$ is well described by the initial settling equilibrium.

We include the values of $\mathcal{R}_{5.2}$ for all runs in Table\,\ref{table:results1}; only a subset are shown in Fig.\,\ref{fig:fig_compare_settling}. Generically, 
a larger opacity gradient results from increasing the ratio $\langle \epsilon^{\rm mid}/\epsilon^{\rm zero} \rangle_{\tilde{\phi}}$, i.e. for stronger settling.
A comparison between both quantities indicates that there is not a simple linear scaling between the opacity gradient and the density gradient, necessitating the full calculation of the opacity, not simply the dust-to-gas ratio, to assess the impact on planetary thermodynamics.

Furthermore, larger values of $\mathcal{R}$ are expected in outer regions of PPDs, where dust scaleheights are reduced.
For instance, if we scale our numerical simulation using the disk model in Section \ref{sec:disk} to $r_p = r_0 = 30 {\rm AU}$, so that $\Sigma_0 \simeq 6 \,{\rm g}\,{\rm cm}^{-3}$ and $T\simeq 27\, {\rm K}$, the mid-plane to pole ratio $\mathcal{R}$ significantly increases.
This estimation is shown in Table\,\ref{table:results1} and denoted by $\mathcal{R}_{30}$.
In all our runs we find that $\mathcal{R}_{30} \gg \mathcal{R}_{5.2}$. 

As shown in Fig \ref{fig:fig1}, the dust distribution is asymmetric both latitudinally and azimuthally. 
To better understand the effect of non-uniform dust on the escaping radiation, we compute shell averages of $1/\bar{\kappa}_{\rm R}$, which is proportional to the optically thick radiation flux. 
To obtain this single average and compare with previous work, including 1D models, we take the integral over a solid angle of the inverse of both opacities, $\bar{\kappa}_{\rm R, mix}$ and $\bar{\kappa}_{\rm R}$.
We define the ratio between the two integrals, $W$:
\begin{equation}
    W = \frac{2\pi}{\bar{\kappa}_{\rm R, mix}} \left( \int^{2\pi}_0 \int^{0}_{\pi/2} \frac{1}{\bar{\kappa}_{\rm R}} \sin(\tilde{\theta}){\rm d}\tilde{\theta} {\rm d}\tilde{\phi} )\right)^{-1}.
\end{equation}

As we have done with $\mathcal{R}$, we estimate $W$ at $r = 5.2 {\rm AU}$ and $r = 30 {\rm AU}$ and show the results in Table\,\ref{table:results1}.
At $5.2 {\rm AU}$, we found that the integrated opacity is lower by a few percent down to 20\%, for the case of the run \texttt{SUB-d6} 
Much smaller values are obtained for $W$ at $r = 30 {\rm AU}$, consistent with the enhanced settling of dust.  
Overall, the integrated inverse opacity decreases with respect to the well-mixed case. Moreover, $W$ decreases with increasing numerical resolution, therefore the obtained values may be taken as an upper bound (see appendix\,\ref{appendix:convergence} for more details).

Moreover, we expect that the results obtained at $r=5.2 {\rm AU}$ are likely conservative in that they underestimate the impact of the flow perturbations introduced by the planet potential at the scales of the Bondi sphere; the deviations from a well mixed distribution increase with increasing numerical resolution (see for example values of runs \texttt{SUB-d5} and \texttt{SUB-d5-half-cnv} and discussion in Appendix\,\ref{appendix:convergence}).

The shell-integrated quantity $W$ may prove useful for benchmarking 1-D evolutionary calculations against  3D multi-fluid simulations with radiative transfer where the opacity is updated based on the dust dynamics. For example, thermal gradients induced by opacity anisotropy are not included in this analysis and are expected to affect the dust distributions. 
We caution that a single value cannot likely capture the full impact of the observed asymmetry.
We stress that the integrated inverse of the opacity by definition favors areas with lower opacity values and therefore may underestimate the impact of the optically thick (mid-plane) regions on cooling.
As we discuss in Section\,\ref{sec:discussion} shell-averaged opacities may not be adequate to compute planetary cooling,  since optical thin and thick regions co-exist for a single planet. Such gradients cannot trivially be cast into a one-dimensional envelope model.

\subsubsection{The adiabatic case}
\label{sec:adia}

So far we have considered only simulations with a locally isothermal equation of state.
However, simulations with an adiabatic equation of state show different gas meridional flow \citep{Kurokawa2018,Fung2019} inside the Bondi sphere. 
To investigate the consistency of our findings with an adiabatic equation of state we expand our parameter exploration to include a simulation where the pressure and internal energy\footnote{Since we consider an adiabatic equation of state we explicitly integrate the internal energy. The initial condition remains as described in Section\,\ref{sec:disk}, although we set $e_{\rm int} = \rho_{\rm g}(hv_{\rm K})^2/(\gamma-1)$. The additional boundary condition is set to $\partial_\theta e = \partial_r e = 0$. We show a comparison with the result from \cite{Fung2019} in Appendix \ref{appendix:convergence}}, $e$, satisfy $P = (\gamma-1)e$. We assume a ratio of the specific heats $\gamma=1.4$, consistent with an ideal diatomic gas.

In the bottom panel of Fig.\,\ref{fig:kappa_average} we show the azimuthal average of the Rosseland and Planck mean opacity for the run \texttt{SUB-half-adia} (adiabatic) and \texttt{SUB-d5}. 
At the scales of the Bondi sphere of the sub-thermal mass planet, there are only minor differences
between the mean opacity gradients.
While we do observe circulation patterns, these do not appear to substantially change the dust distribution as a function of height. Since the latitudinal gradient is primarily driven by the initial settling equilibrium, the deviations in the opacity compared to the well-mixed case persist.

We emphasize that because we adopt an initial condition for this run that is in equilibrium for an isothermal equation of state longer integration times may show different results.  
Nevertheless, we expect irradiated disks in the regime of interest to be closer to the isothermal regime than adiabatic (due to efficient cooling), even if the gas within the planetary envelope undergoes adiabatic perturbations. 
A full radiative transfer treatment will be necessary to explore more realistic planetary cooling, and is the subject of future work.

\subsection{Optically thin and thick regimes}
\label{sec:discussion}

The results described in Section\,\ref{sec:results_opacity} suggest that the latitudinal opacity gradient is driven by dust settling,  and modestly steepened by the planet potential. 
Therefore, to first order we can determine conditions that lead to this opacity gradient assuming axisymmetric disk models, with different size-distributions and dust scaleheights. 
Moreover, to better  estimate  the  impact  of this  opacity  gradient on the planets radiative processes we calculate the photon mean-free-path defined as $\lambda_{\rm mfp}  \equiv 1/(\rho_{\rm g}\bar{\kappa}_{\rm P})$.
We adopt a cylindrical coordinate system $R, Z$, and  assume a settling-diffusion equilibrium dust distribution where the density associated with a given size $a_k$ is obtained as
\begin{align}
    \label{eq:dust_eq}
    \rho_{{\rm d}, k} &= \rho_{\rm g} \frac{\epsilon(a_k)}{h_{\rm d,k}}  \exp\left( - \frac{T_{{\rm s}k}}{\delta}\left[ e^{\frac{Z^2}{2H^2}} -1\right] \right) \\
     \rho_{\rm g} &= \frac{\Sigma_0}{\sqrt{2\pi}H} R^{-\sigma}_{\rm au}  e^{-\frac{Z^2}{2H^2}}  \nonumber 
\end{align}
where $R_{\rm au}$ corresponds to the distance to the central star in astronomical units. 
The gas scale-height is denoted as $H = h(R) R$ with $h(R) = h_0 R^f_{\rm au}$, and the Stokes number is defined as $T_{{\rm s}k} = \pi a_k \rho_{\rm solid} /(4\Sigma_0 R^{-\sigma}_{\rm au})$ \citep[see e.g.,][]{Dipierro2018}. 
The values of $\epsilon(a_k)$ and $h_{{\rm d }k}$ are obtained replacing from Eq.\,\ref{eq:distribution} and Eq.\,\ref{eq:h_d}, respectively. 
We  focus on scales of the Bondi radius of sub-thermal mass planets in disks with moderate-to-strong dust settling. 

\begin{figure*}[t]
    \centering
    \includegraphics[scale=0.82]{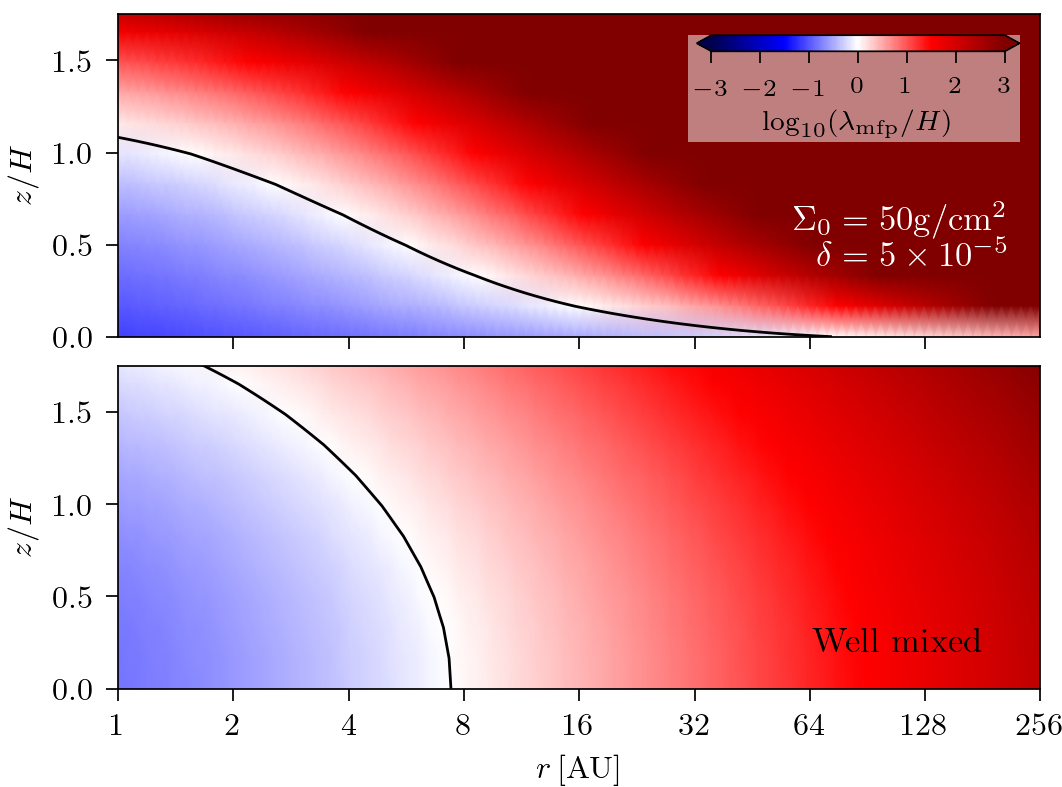}\hfill
    \includegraphics[scale=0.82]{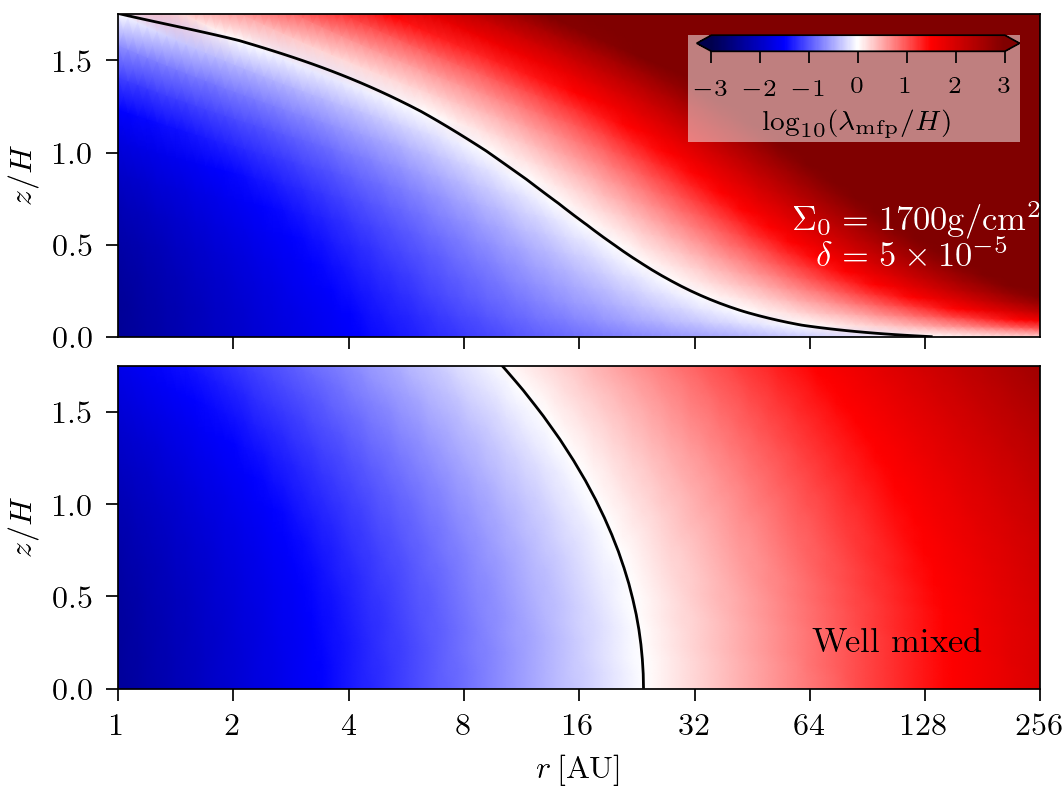} \\
    \includegraphics[scale=0.82]{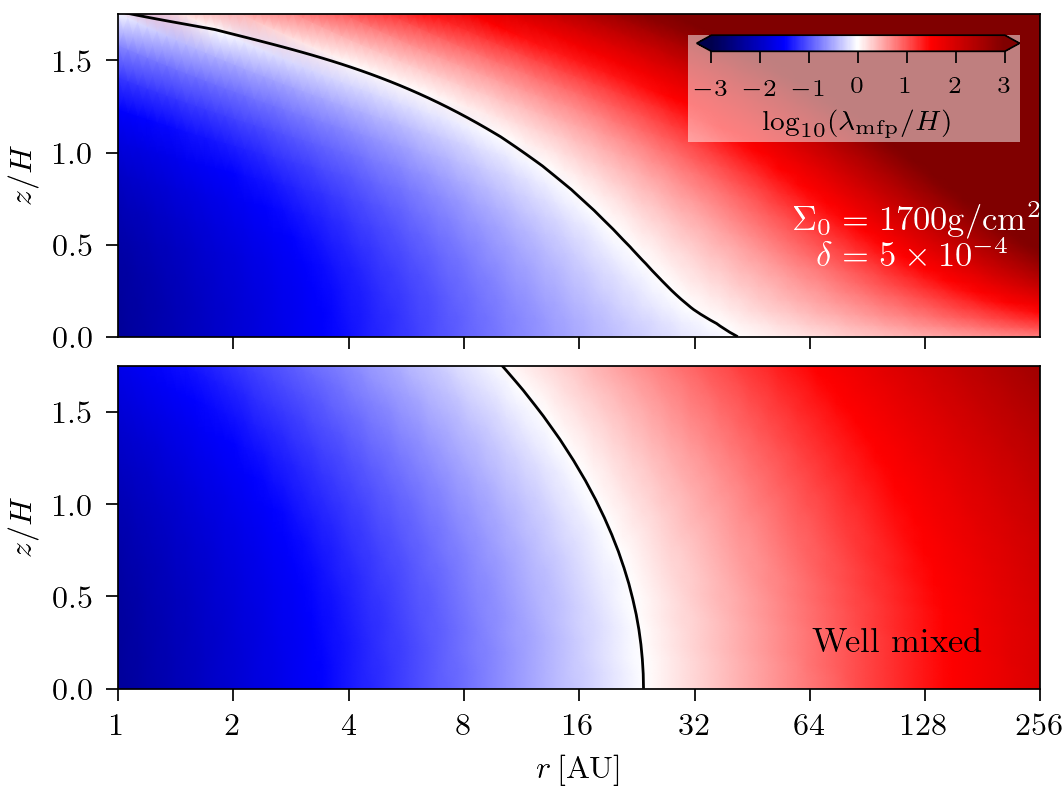}\hfill
    \includegraphics[scale=0.82]{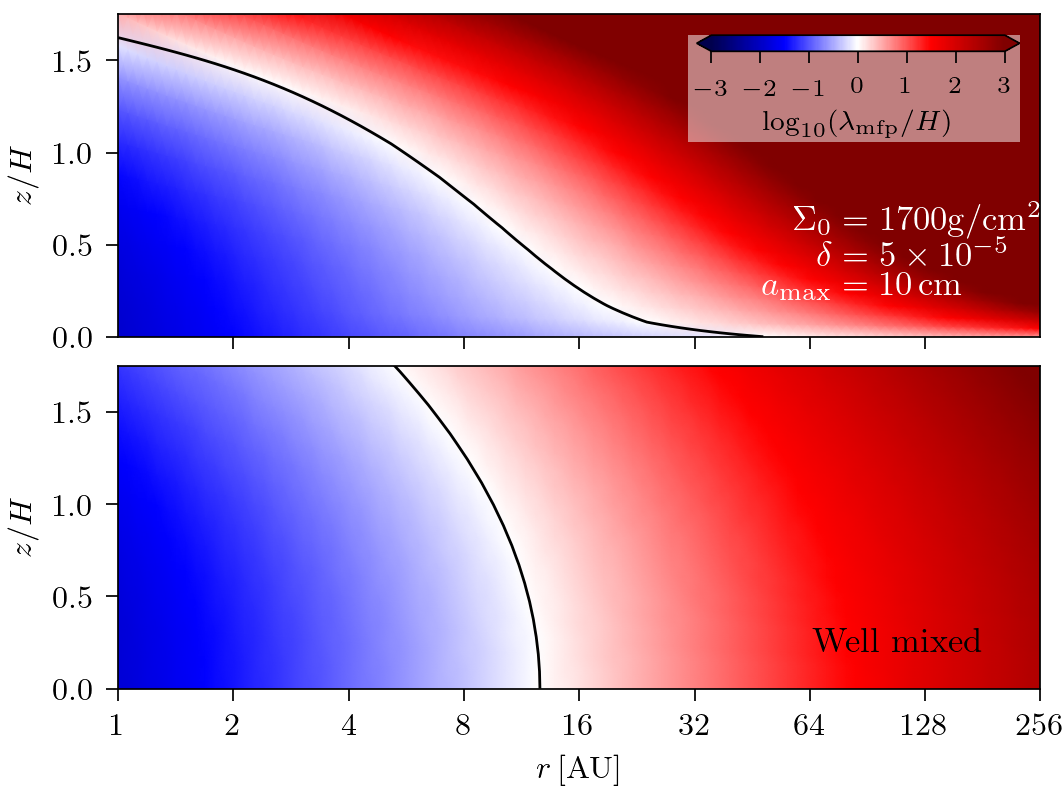}
    \caption{Photon mean-free-path, $\lambda_{\rm mfp} \equiv 1/(\bar{\kappa}_{\rm P} \rho_{\rm g})$ for different disk models with initial dust settling profiles and size distributions.
    The minimum particle size is fixed to $1\mu {\rm m}$ and the maximum particle size is assumed $a_{\rm max}=1 {\rm cm}$ unless specified. 
    Except for the top left panel where the gas density and scale-height were adopted from the disk model used for the run \texttt{SUB-d5-sig0}, the rest of the results are obtained from a MMSN disk model with $h_0=0.033$, $f=0.25$ and $\sigma=3/2$ (see Eq.\,\ref{eq:dust_eq}).
    Our estimations suggest that planet envelopes may have a transition between optically thin and optically thick regimes at the Bondi radius in outer regions of the disk. The radial location and planet mass where the transition occurs are a function of the adopted dust size-distribution, the gas surface density, disk temperature, and flaring index. }
    \label{fig:fig_mfp}
\end{figure*}

We adopt a criterion for transition between optically thick and thin regimes when $\lambda_{\rm mpf} = H$. 
Note that this is consistent with a criterion given by $\lambda_{\rm mpf} = r_{\rm B}$ \citep[e.g.,][]{Rafikov2006} for planets with nearly one thermal mass.

In Fig.\,\ref{fig:fig_mfp} we show  $\lambda_{\rm mfp}$ for different disk models, with variable gas surface density, diffusion parameters, and dust properties,  assuming the opacity gradient is set soley by dust settling.
The top left panel corresponds to a case similar to \texttt{SUB-d5-sig0}, while the others were obtained for a Minimum Mass Solar Nebula (MMSN) disk model \citep{Hayashi}.
Note that the MMSN models are flared, rather than constant aspect ratio. For all disk models, the well-mixed distribution of solids has a transition between optically thin and optically thick regimes that is mainly a function of the radial distance to the central star for $Z\lesssim H$.

\begin{figure}[t]
    \centering
    \includegraphics[]{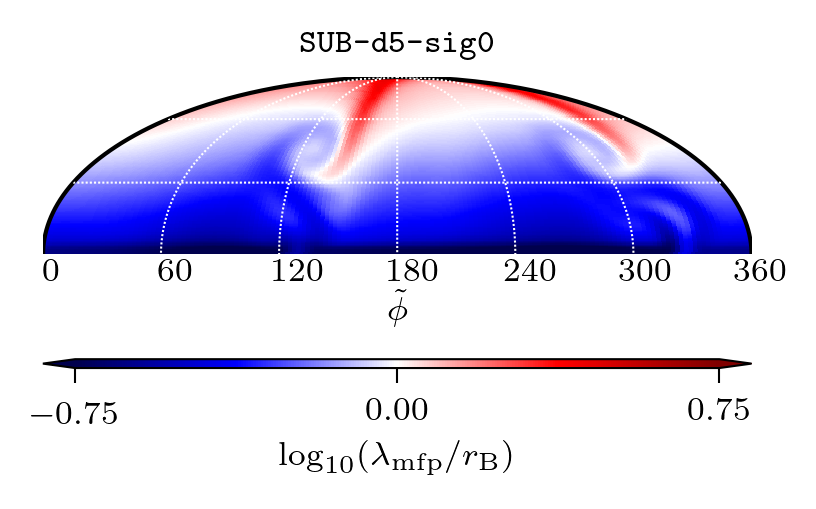}
    \caption{Photon mean-free-path, $\lambda_{\rm mfp} \simeq 1/(\bar{\kappa}_{\rm P} \rho_{\rm g})$ at the Bondi semi-sphere for the run \texttt{SUB-d5-sig0}  at $r=5.2 \rm{AU}$.
    The transition between optically thin and optically thick regimes is in agreement with the results of top left panel from Fig.\,\ref{fig:fig_mfp} obtained from the settling equilibrium condition with no planet.
    }
    \label{fig:fig_mfp_sig0}
\end{figure}

On the other hand, when accounting for settling, we find a shallower transition between the two regimes as a function of radius. Therefore the envelopes of sub-thermal mass planets may be optically thick near the mid-plane and optically thin near the polar regions.
This transition is confirmed for our run \texttt{SUB-d5-sigma0} (see Fig.\,\ref{fig:fig_mfp_sig0}) where the planet is assumed to be at $r=5.2{\rm AU}$ from the central star. 

As can be appreciated from Fig.\,\ref{fig:fig_mfp}, the transition near $\lambda_{\rm mfp} \sim H$ is more likely to occur at intermediate regions of PPDs, whereas inner regions are optically thick and outer regions are optically thin. 
At a distance from the star $r \lesssim 10 \rm{AU}$, such a transition would likely occur in nearly laminar and/or gas depleted disks, as the required settling-diffusion equilibrium parameter should be $\delta < 10^{-5}$.

We emphasize that the estimates of $\lambda_{\rm mfp}$ strongly depend on the size distribution and the adopted disk model. 
In particular, the gas surface density and the diffusion parameter will set the equilibrium scale height of dust grains.
In addition, the maximum grain size and the slope of particle-size distribution also impact the mean free path transition as they set the mass load towards the micron sizes grains for a fixed dust-to-gas ratio.

The results obtained in Fig.\,\ref{fig:fig_mfp} and Fig.\,\ref{fig:fig_mfp_sig0} suggest that some planets may occupy a regime not captured by 1D models that are either purely thick ($\lambda_{\rm mfp} \ll r_{\rm B}$) or thin regimes ($\lambda_{\rm mfp} \gg r_{\rm B}$), as studied in \cite{Rafikov2006}; 
Future 1D models may benefit from seeking an approximation that captures the transition of $\lambda_{\rm mfp}$ reported in this work.

Similar axisymmetric dust opacity calculations have been used to study thermal relaxation in PPDs including the exchange of energy between gas and dust via collisions \citep[][]{Malygin2017,Barranco2018,Bae2021}, and self-consistent coagulation models \citep[][]{Sengupta2019,Chachan2021}.  Ultimately, whether the observed latitudinal gradient plays a significant role in the thermodynamic evolution of planets requires properly coupling the planet radiation, dust distribution, and disk hydrodynamics.

\section{Conclusions}\label{sec:conclusions}

In this work, we have performed three-dimensional multi-fluid simulations of nearly thermal mass planets embedded in a protoplanetary disk. 
Our simulations, applied to a standard disk model with a planet at 5.2(30) AU, allowed us to characterize a dust-distribution with particle sizes spanning from $\sim 1\, \mu {\rm m}$ ($\lesssim 0.1 \mu {\rm m}$) to $\sim 1\, {\rm cm}$ ($\lesssim 0.05\, {\rm cm}$) at the Bondi sphere of a $8.7$ Earth-mass planet and at the Hill sphere of a $26$ Earth-mass planet \footnote{Since we assume a constant aspect ratio in our simulated disks, the planet thermal masses considered at $5.2$ and $30$ AU are the same.}. 
While we have included dust feedback in all our runs, a comparison with gas-only simulations confirmed the overall planet atmospheric structure is insensitive to the dust drag-force, at least for dust-to-gas mass ratios of $\lesssim 0.1$. 

We found that the dust equilibrium scale-height plays a primordial role in setting the much stronger latitudinal gradient of the (azimuthally averaged) opacities.
The planet's gravitational force only modestly steepens the latitudinal gradient at the scales of the Bondi and Hill semi-spheres. 

Our numerical simulations revealed the presence of a persistent anisotropy in the dust opacities on the relevant scales of $r_{\rm B}$ and $r_{\rm H}$.
The anisotropy is established after two planetary orbits and remains in a steady-state up to $80$ orbits. 
While smaller grains have even longer settling times, they are more well-mixed and do not significantly participate in the anisotropy.
We find that spiral wakes from the planet generically introduce azimuthal dust asymmetries near the mid-plane.

Our results indicate that planetary envelopes may have a latitudinal transition between optically thin and optically thick regimes, which may limit the applicability of 1D models where such a transition cannot be captured. 
This transition is likely to occur when the scale-height of grains that set the peak wavelength of a black body radiation is smaller than the planet Bondi radius. 
At temperatures between $10\,$K to $100\,$K, the peak wavelength correspond to grains with sizes smaller than $100 \mu {\rm m}$.
Hence we expect that disks with lower surface densities and/or weaker gas stirring  will show this transition.
We found that for standard MMSN surface density and settling parameters $\delta \lesssim 5 \times 10^{-4}$ the transition between optically thick and thin regimes occurs for nearly thermal mass planets at $r\gtrsim 10 {\rm AU}$. 

A latitudinal transition in the mean-free-path differs from the traditional, symmetric case where envelopes are either optically thick or thin \citep[see e.g.,][]{Rafikov2006,Lee2015}.
Moreover, the three-dimensional dust-dynamics produces an opacity that can deviate from opacity calculations based on the results from \cite{Bell1994} and \cite{Semenov2003}. 
The integrated opacity, and consequently the cooling time, is reduced in comparison with that obtained from a well-mixed size distribution (see for example cases with $W_{30}$), which could favor the faster growth of massive planets \citep[][]{Hubickyj2005, MOVSHOVITZ2010}.
Note that our opacity calculations for a planet at $r=30 {\rm AU}$ show even larger opacity gradients, and smaller integrated opacities, implying a greater impact on planets forming in the outer regions of PPDs.

To better understand the scope of our results, future simulations should include turbulence and/or winds as a mechanism of stirring dust, rather than a standard diffusion flux.
We, however, stress that the assumed dust equilibrium scale-height are in reasonable agreement with regimes dominated by Magneto Rotational Instability (MRI) turbulence \citep[][]{Flock2017} and MHD winds at $r \gtrsim 30 \rm{AU}$ \citep[e.g.,][]{Riols2018}.
Moderate-to-strong settling  of dust grains in outer regions of PPD has been also inferred from observations of HL Tau \citep[][]{Pinte2016}.
Note that this is not necessarily the case for self-consistent stirring driven by turbulence as a byproduct of the Vertical Shear Instability (VSI) \citep[see e.g,][]{Stoll2016, Picogna2018, Flock2017}.

Ultimately, three dimensional multi-fluid simulations coupled with self-consistent radiative transfer are required to fully address the scope of our results.
In addition to impacts on cooling and accretion rates, we anticipate that our results may also have implications for thermal physics of the heating torque \citep[][]{Benitez-Llambay2015}, the evolution of eccentricity and inclination of hot protoplanets \citep[e.g.,][]{Eklund2017,Chrenko2017} and the excitation of buoyancy resonances \citep[e.g.,][]{Zhu2012, Lubow2014, McNally2020, Bae2021}.

\section{Software and third party data repository citations} 

All the data was generated with a modified version of the open-source software FARGO3D available at https://bitbucket.org/fargo3d/public.git. The data underlying this article will be shared on reasonable request to the corresponding author.

\acknowledgments
We thank Pablo Ben\'itez-Llambay for useful suggestions and valuable contributions.
We thank Phil Armitage and Zhaohuan Zhu for inspiring discussions that motivated this work.
Finally we thank the referee for the thorough report. We acknowledge support from Grant 80NSSC19K0639 and useful discussions with members of the TCAN collaboration.  Numerical simulations were powered by the El Gato supercomputer supported by the National Science Foundation under Grant No. 1228509.  An allocation of computer time from the UA Research Computing High Performance Computing (HPC) at the University of Arizona is gratefully acknowledged

\appendix

\section{Convergence with numerical resolution} \label{appendix:convergence}

To validate our results, we perform a resolution study varying the number of grid cells across the Bondi radius and the extent of the azimuthal domain. We also benchmark our results against both isothermal and adiabatic simulations from the literature with planet mass  $M_p \simeq 0.6 M_{\rm th}$.
Referencing table \ref{tab:runs}, we utilize  runs \texttt{SUB-d5}, \texttt{SUB-d5-half}, \texttt{SUB-d5-half-cnv} and \texttt{SUB-d5-half-adia}.

We focus our literature comparisons on \cite{Fung2019}, who performed 3D global simulations to study the formation of circumplanetary disks, employing an isothermal and an adiabatic equation of state.
They found that envelopes of sub-thermal mass planets are nearly in hydrostatic equilibrium with little rotational support at the scales of the Bondi radius. 
Neglecting the stellar potential, the hydrostatic equilibrium around the planet gives a density profile that follows as
\begin{equation}
    \label{eq:iso}
    \rho_{\rm g} = \rho_0 \exp\left( \frac{r_{\rm B}}{\sqrt{r^2 + r^2_{\rm s}}} \right)\,,
\end{equation}
for the isothermal case, whereas for the adiabatic equation of state (assuming an isentropic process) the density profile is
\begin{equation}
    \label{eq:adi}
    \rho_{\rm g} = \rho_0 \left( 1 - \left(\frac{1}{\gamma} - 1\right)\frac{r_{\rm B}}{\sqrt{r^2 + r^2_{\rm s}}} \right)^{\frac{1}{\gamma-1}}\,,
\end{equation}
where $\gamma=1.4$ is the adiabatic index and $r_{\rm s}$ is the smoothing length (for the runs \texttt{SUB-d5-half-cnv} and \texttt{SUB-d5-half-adia} $r_{\rm s}$ is about six percent of the Bondi radius).

\begin{figure*}[h]
    \includegraphics[scale=0.65]{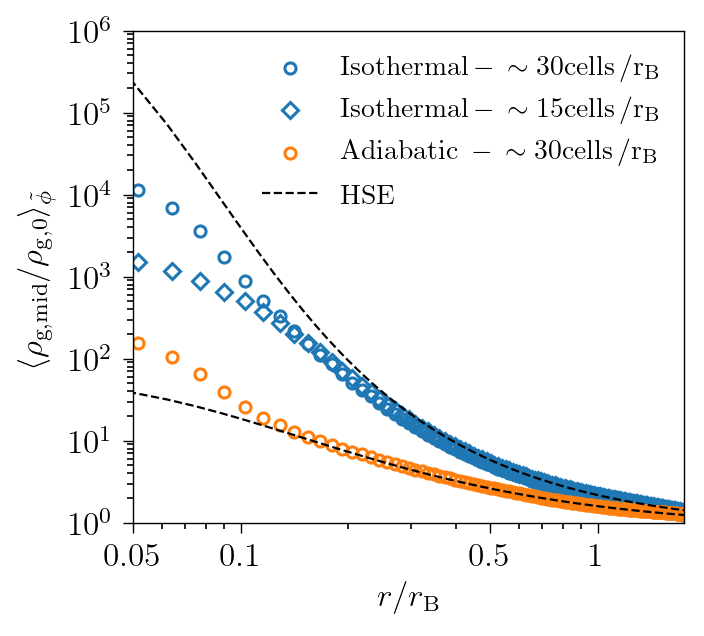}
    \includegraphics[scale=0.65]{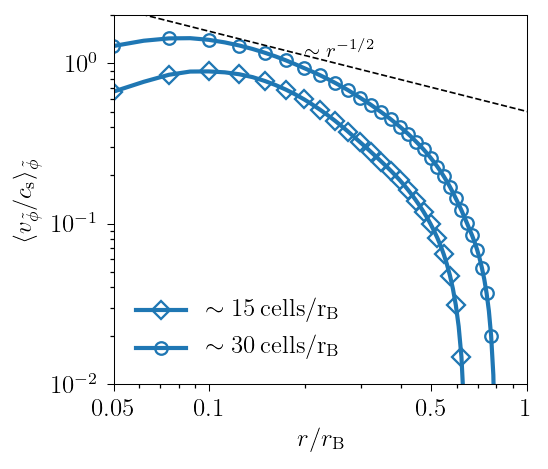}
    \includegraphics[scale=0.675]{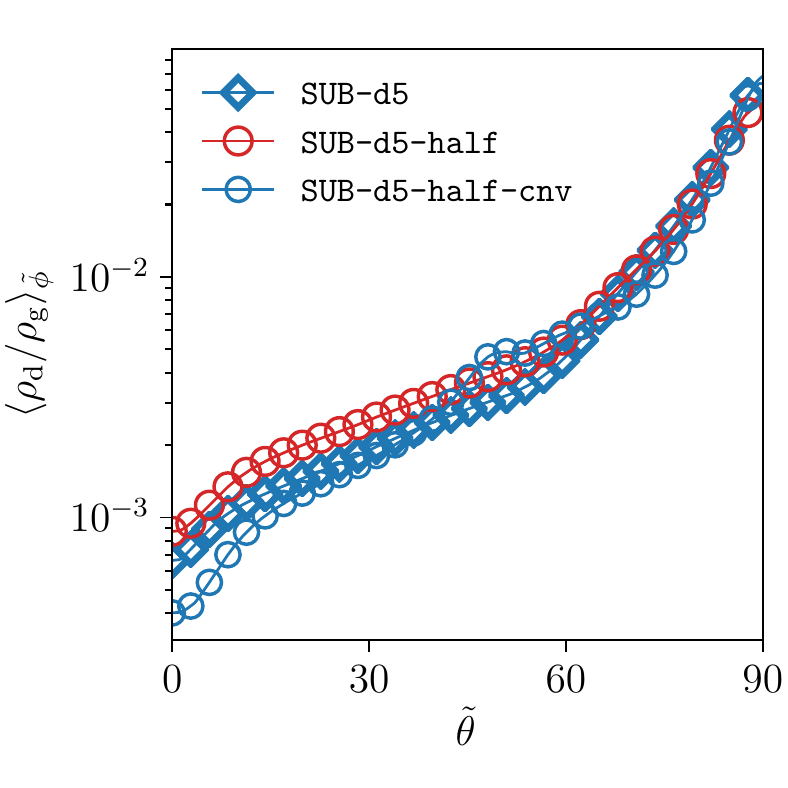}
    \caption{Convergence with resolution. 
    Left panel: Azimuthal-average of the gas density at the mid-plane for the runs \texttt{SUB-d5} ($15 \rm{cells}/r_{\rm B}$), \texttt{SUB-d5-half-cnv} ($30 \rm{cells}/r_{\rm B}$) and \texttt{SUB-d5-half-adia} ($30 \rm{cells}/r_{\rm B}$). In the three cases the density profile is in decent agreement with the hydrostatic equilibrium obtained considering only the softened planet potential. 
    Center panel: Azimuthal-average of the gas azimuthal velocity in a cylindrical coordinate system centered at the planet location for the runs \texttt{SUB-d5} ($15 \rm{cells}/r_{\rm B}$) and \texttt{SUB-d5-half-cnv} ($30 \rm{cells}/r_{\rm B}$). In agreement with the strong pressure support at the Bondi radius (left panel) the rotational support is negligible as $r \rightarrow r_{\rm B}$. The inner regions partially approach a Keplerian profile. Although these regions are poorly resolved, the results are in reasonable agreement with those from \cite{Fung2019} (see Figure). 
    Right panel: Azimuthal average of the total dust-to-gas mass ratio at the Hill semi-sphere for the runs \texttt{SUB-d5} ($15 \rm{cells}/r_{\rm B}$), \texttt{SUB-d5-half} ($15 \rm{cells}/r_{\rm B}$ and half azimuthal domain) and \texttt{SUB-d5-half-cnv} ($30 \rm{cells}/r_{\rm B}$ and half azimuthal domain.  )   }
    \label{fig:fig_appendix_a}
\end{figure*}

 In the left panel of Fig.\,\ref{fig:fig_appendix_a} we compare the radial profile for the runs \texttt{SUB-d5}, \texttt{SUB-d5-half-cnv} and \texttt{SUB-d5-half-adia} with the solutions given in Eqs.\,\ref{eq:iso} and \ref{eq:adi}.
In all cases the numerical solutions are in good agreement with the hydrostatic equilibrium solution at $r/r_{\rm B}\gtrsim 0.2$. The differences between the density profiles at $r/r_{\rm B}<0.2$ are driven by the choice of smoothing length, which is set to 2 cells in azimuth, 2.3 cells in radius and 3 cells in latitude.  
 
In the middle panels of Fig.\,\ref{fig:fig_appendix_a}, we show the radial profile of the azimuthal velocity in a cylindrical coordinate system centered on the planet.
The profiles were calculated in the mid-plane after taking the azimuthal average on a cylindrical mesh centered on the planet.
Pressure support dominates from the outermost regions of the Bondi sphere down to $r \sim r_{\rm B}/2$. 
These results are in agreement with previous work showing increasing rotational support closer to the planet, where circumplanetary disks can for \citep{Fung2019}. Our resolution is insufficient to identify any Keplerian rotation below  $r \sim r_{\rm B}/2$.

In the rightmost panel of Fig.\,\ref{fig:fig_appendix_a} we show the azimuthal average of the total dust-to-gas density ratio at the Bondi sphere for the runs \texttt{SUB-d5} (full azimuthal domain), \texttt{SUB-d5-half} and \texttt{SUB-half-cnv} (half azimuthal domain). 
The mean dust-to-gas density ratio is in good agreement for the runs \texttt{SUB-d5} and \texttt{SUB-d5-half}, indicating that the choice of a half-domain in azimuth may not affect the overall dust-density and therefore opacity distribution at the scales of the Bondi sphere.   
\begin{figure*}[h]
    \centering
    \includegraphics[scale=0.8]{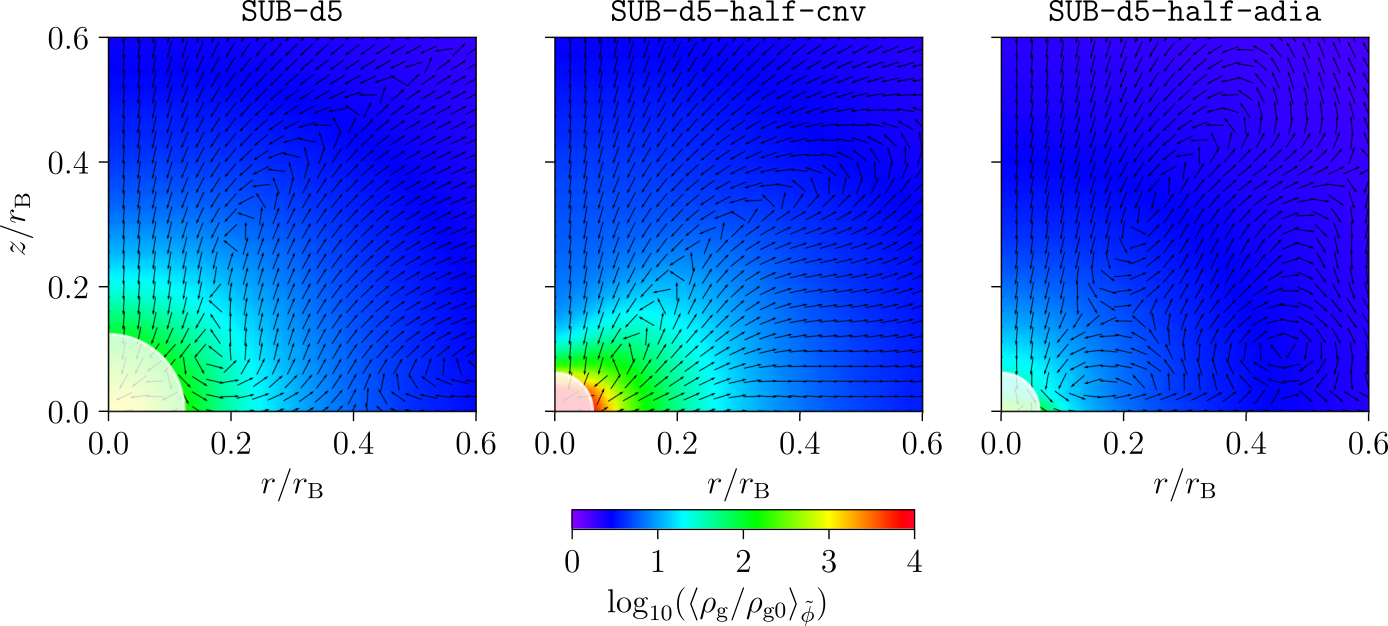}
    \caption{Azimuthal-average of the gas density and meridional velocity field for the runs \texttt{SUB},  \texttt{SUB-half-cnv} and \texttt{SUB-ad}. The left and center panels correspond to the isothermal case but with different resolutions, whereas the results from the adiabatic are displayed in the rightmost panel.     The meridional flows show a recycling pattern with  polar inflow and mid-plane out-flow. While these simulations include dust, the dust-feedback seems negligible and therefore these results may be compared with those from \cite{Fung2019}. The important discrepancies between the left and center panel highlight the inadequate resolution of our fiducial run below $r=r_{\rm B}/2$.  }
    \label{fig:fig_appendix}
\end{figure*}

At scales of $r_B$, the runs have not entirely converged in the polar regions. We found that the dust-to-gas density ratio is more depleted at regions $\tilde{\theta} \sim 0$ for the case \texttt{SUB-half-cnv}.
This has an impact on the opacity gradient as mentioned in Section\,\ref{sec:results_opacity}. 
More precisely,  we found that the integrated inverse opacity is $\sim7\%$  smaller  than  the  well-mixed  case for the run \texttt{SUB-d5}, whereas the largest resolution gives a discrepancy of $\sim22\%$. Thus we expect that our results are conservative, in that we have likely underestimated the magnitude of the opacity gradient.

Finally, in Fig.\,\ref{fig:fig_appendix}, we show the azimuthal average of the density along with the meridional velocity field for the same runs displayed in Fig.\,\ref{fig:fig_appendix_a}.
Our results are in good agreement with those described in \citep[][]{Fung2019}. Note that the detailed behavior of the flow at $r< r_{\rm H}/2$ may not be accurate given the simulations typically have $\lesssim 8$ cells at that radius, except for  runs $\verb|SUB-half-cnv|$ and $\verb|SUB-half-adia|$ where we have $\lesssim 15$ cells at half of the Bondi radius. 
Our comparison suggest that a minimum of $30 {\rm cells} / r_{\rm B}$ is required to characterize the meridional circulation pattern and dust density distribution below half of the Bondi radius. 
Moreover, for the explored parameters and timescales (less than 100 orbits) we found that simulations with reduced azimuthal domain may help to expedite the calculations without significantly affecting the dynamics near the planet. 
%


\end{document}